\documentclass[10pt,a4paper,twocolumn]{article}
\usepackage[a4paper,left=16mm,right=16mm,bottom=25mm,top=25mm]{geometry}
\usepackage[switch]{lineno} 
\usepackage{lipsum} 
\usepackage{footmisc}  
\usepackage[fleqn]{amsmath}
\usepackage{fancyhdr}
\usepackage{subfigure}
\usepackage{graphicx}
\usepackage{booktabs}
\usepackage[british]{babel}
\usepackage[square,comma,numbers,sort&compress]{natbib}
\usepackage{comment}
\usepackage{lettrine}

\usepackage{csvsimple}
\usepackage{graphicx}
\usepackage{subcaption}
\usepackage{amsmath}
\usepackage{tikz}
\usepackage{array}
\usepackage{tabu}
\usepackage{multirow}
\usepackage{url}
\usepackage{xcolor}
\usepackage{afterpage}
\usepackage{graphicx}%
\usepackage{bm}%
\usepackage[title]{appendix}
\usepackage[]{subfigure}
\usepackage{float}
\usepackage{hyperref}
\hypersetup{colorlinks,linkcolor={blue},citecolor={blue},urlcolor={cyan}}
\usepackage{cleveref}
\usepackage{afterpage}
\usepackage{caption}
\usepackage{lineno}
\usepackage{lettrine}
\usepackage{color}
\usepackage{soul}
\usepackage{ulem}
\usepackage{float}
\usepackage{placeins}

\newcommand{\bluemark}[1]{\textcolor{black}{#1}}

\definecolor{greendark}{rgb}{0.0,0.5,0.0}
\definecolor{darkred}{rgb}{0.55, 0.0, 0.0}
\definecolor{navy}{rgb}{0.0, 0.0, 0.5}
\usepackage{graphicx}

\title{Hybrid d/p-wave altermagnetism in Ca$_{3}$Ru$_{2}$O$_{7}$ and strain-controlled spin splitting}

\author{Andrea León$^{1,2,*}$, Carmine Autieri$^{3}$, Thomas Brumme$^{4}$, Jhon W. González$^{5,\dag}$}
\date{
$^{1}$Departamento de Física, Facultad de Ciencias, Universidad de Chile, Casilla 653, Santiago, Chile.\\
$^{2}$Dresden University of Technology, Institute for Solid State and Materials Physics, 01062 Dresden, Germany \\
$^{3}$International Research Centre Magtop, Institute of Physics, Polish Academy of Sciences, Aleja Lotników 32/46, 02668 Warsaw, Poland.\\
$^{4}$Chair of Theoretical Chemistry, Technische Universität Dresden, Bergstrasse 66, 01069 Dresden, Germany.\\
$^{5}$Departamento de Física, Universidad de Antofagasta, Av. Angamos 601, Casilla 170, Antofagasta, Chile.\\[2ex]
}

\begin{document}
\twocolumn[
\begin{@twocolumnfalse}
\maketitle

\vspace{10pt} 
\noindent
\begin{small} 
\setlength{\parskip}{0pt} 
\setlength{\parindent}{0pt} 
\noindent The interplay of strong electronic correlations, sizable octahedral distortions, and pronounced spin–orbit coupling (SOC) makes perovskite oxides promising candidates for realizing altermagnetic phases.  We study altermagnetic phases in Ca$_3$Ru$_2$O$_7$, a non-centrosymmetric layered perovskite whose ground state is a Kramers-degenerate antiferromagnet.  We show that an alternative Néel-type spin arrangement hosts a P-2 $d$-wave altermagnetic state with orbital selectivity similar to Ca$_2$RuO$_4$.  Including SOC generates a symmetry-allowed $p$-wave component and yields a hybrid $d$/$p$-wave altermagnetic order.  We further demonstrate that biaxial strain tunes both magnetic stability and band splitting: compressive strain beyond 2\% favors the altermagnetic phase over the antiferromagnetic ground state, while tensile strain increases altermagnetic splittings by up to 9\%.  To quantify these trends, we define an altermagnetic figure of merit and trace its strain dependence to changes in electronic localization and octahedral geometry in this polar metal.\\

\end{small}
\vspace{15pt} %

\end{@twocolumnfalse}]

\section*{Introduction}
\footnotetext[1]{\textsuperscript{*} Corresponding author: andrea.leon@uchile.cl}
\footnotetext[2]{\textsuperscript{\dag} jhon.gonzalez@uantof.cl}

\lettrine[lines=2, loversize=0.25, findent=2pt]{\LettrineFont{\bluemark{A}}}
ltermagnets are a significant new magnetic material class, advancing fundamental research and spintronic applications \cite{vsmejkal2022beyond,bai2024altermagnetism}. At the heart of altermagnetism (AM) lies crystal symmetry, which dictates the unique spin-splitting behavior. This inherent link makes strain a powerful tool for manipulating and controlling AM order in correlated materials, a frontier of active research \cite{sato2024altermagnetic,PhysRevB.109.144421}. The potential of strain engineering has been recently demonstrated by the induced phase transition from an antiferromagnetic to an altermagnetic state in ReO$_2$ \cite{PhysRevB.109.144421} and the discovery of an elasto-Hall conductivity in d-wave altermagnets \cite{takahashi2025elasto}. The physics is further enriched by spin-orbit coupling (SOC), which, by breaking time-reversal symmetry, can induce phenomena such as weak ferromagnetism through spin canting or other mechanisms dependent on crystal symmetry \cite{DZYALOSHINSKY1958241,autieri2024staggereddzyaloshinskiimoriyainducingweak,roig2024quasisymmetryconstrainedspinferromagnetism,cheong2025altermagnetismclassification,Cheong2024,PhysRevB.110.155201,jo2025weakferromagnetismaltermagnetsalternating}. Collectively, these findings motivate the exploration of strain-induced AM order in new classes of correlated materials. In this context, Ruddlesden-Popper perovskites, renowned for hosting diverse quantum phases intrinsically coupled to lattice distortions, offer a compelling platform to investigate the interplay between strain, symmetry, and magnetism \cite{bernardini2025ruddlesden,autieri2025conditionsorbitalselectivealtermagnetismsr2ruo4,naka2025altermagnetic,bernardini2024ruddlesden,li2024dwavemagnetismcupratesoxygen}.

\bluemark{Ruddlesden-Popper compounds of the form (Ca, Sr)$_{n+1}$Ru$_n$O$_{3n+1}$ exhibit strong electronic correlations and display phases that depend critically on symmetry and octahedral distortions. Rotations, Jahn-teller distortions, and polar displacements break the operations that interchange opposite spin sublattices and enable the emergence of altermagnetic states \cite{bernardini2024ruddlesden}. Over the past decade, researchers have observed Mott insulating behavior \cite{ricco2018situ}, metamagnetism \cite{tiwari2023suppression}, and unconventional superconductivity \cite{maeno2024still} in these systems. Revisiting their magnetic phases within the altermagnetic framework offers new insights into transitions among ferromagnetism, antiferromagnetism, and altermagnetism by linking symmetry, distortion, and magnetic order. Among the series, Ca$_3$Ru$_2$O$_7$ undergoes antiferromagnetic transitions as a function of temperature \cite{yoshida2005crystal,kikugawa2007ca3ru2o7,PhysRevB.102.235104} and exhibits pronounced tunability under magnetic field \cite{ohmichi2004colossal}, pressure \cite{karpus2006spectroscopic} and strain \cite{dashwood2023strain}. These observations raise the question of whether Ca$_3$Ru$_2$O$_7$ (hereafter CRO) hosts an altermagnetic phase and what conditions might induce a transition from its native antiferromagnetic order.}

\bluemark{We apply this framework to Ca$_3$Ru$_2$O$_7$, a non-centrosymmetric compound with strong spin-orbit coupling (SOC) that enables relativistic spin splitting and spin-momentum locking. SOC governs the magnetic anisotropy in this system and drives spin reorientation transitions, as previously reported \cite{markovic2020electronically}.
 Investigating altermagnetism in Ca$_3$Ru$_2$O$_7$ offers a unique opportunity to reveal spin textures that emerge from the interplay of non-relativistic and relativistic effects. Most known altermagnets preserve inversion symmetry; finding altermagnetic order in a non-centrosymmetric metal would open new directions in the study of symmetry-protected spin transport phenomena \cite{smejkal2024altermagneticmultiferroicsaltermagnetoelectriceffect,D3NR04798A}.}

In this work, we identify magnetic configurations in Ca$_3$Ru$_2$O$_7$ that host altermagnetic order and analyze their symmetry and electronic structure under non-relativistic and relativistic spin splitting. In the absence of SOC, the system exhibits a $d$-wave altermagnetic state, while SOC introduces additional asymmetries and generates a hybrid $d$- and $p$-wave character. We demonstrate that biaxial strain tunes the altermagnetic response by enhancing or suppressing spin splitting depending on the strain direction and link this manipulation to changes in octahedral distortions and orbital localization. To quantify these effects, we define the altermagnetic quantity (AMQ) as a figure of merit that captures changes in band splitting and enables direct comparison across altermagnetic systems across different conditions.

\section*{Results}
\subsection*{Exploring AM states}

\begin{figure}[]
\includegraphics[width=0.5\textwidth]{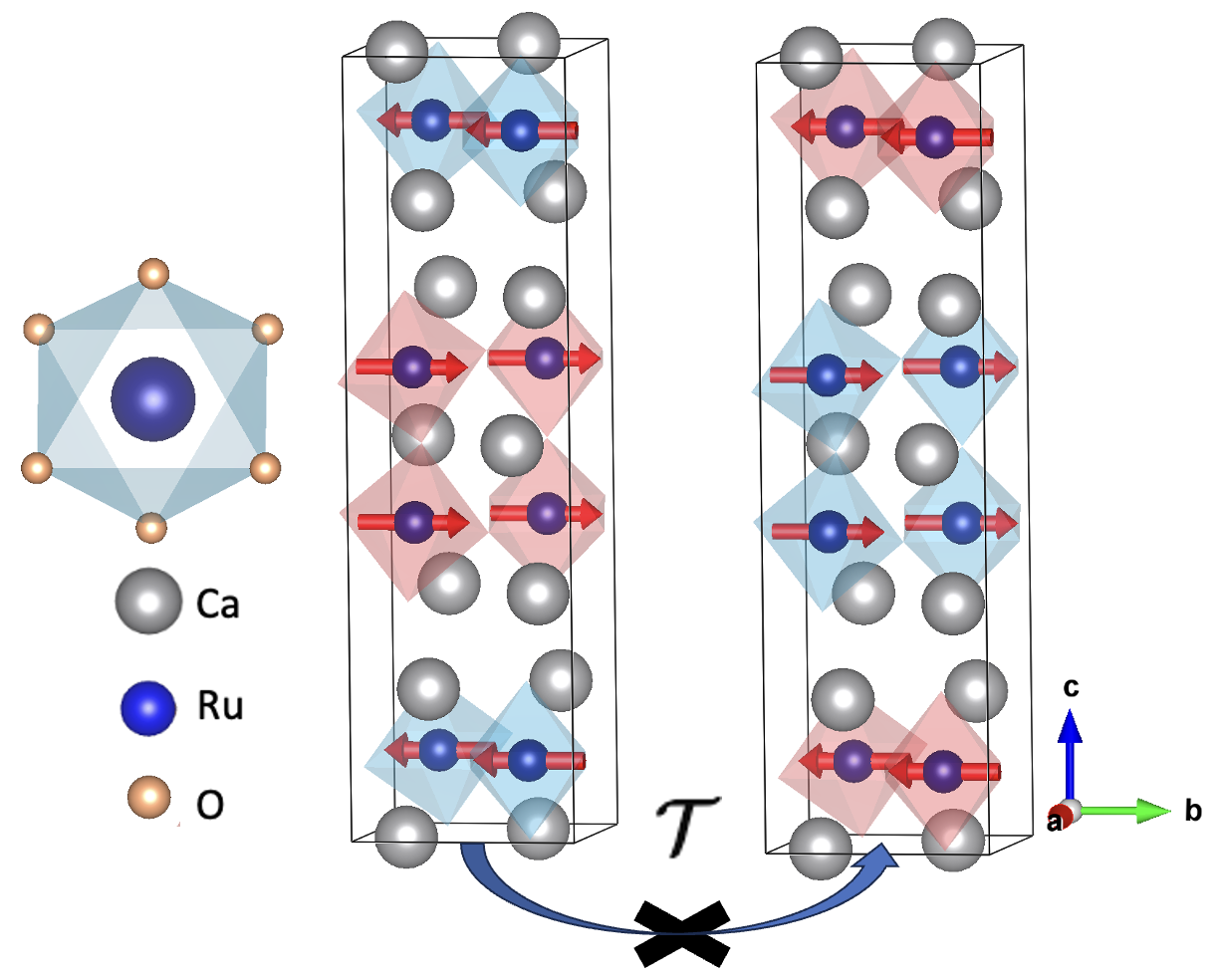}
\caption{Ca$_{3}$Ru$_{2}$O$_{7}$ in its magnetic ground-state configuration exhibits spins aligned ferromagnetically (FM) within the plane and antiferromagnetically (AFM) between layers. This schematic illustrates that, under time-reversal operation, the spins cannot be connected by rotational symmetry alone because they can be connected through translational symmetry operations along the c-axis. These Kramers antiferromagnets are also named SST-3\cite{Yuan2023} in the literature.
Consequently, Ca$_{3}$Ru$_{2}$O$_{7}$ exhibits Kramers antiferromagnetism in its ground state.
}
\label{fig0}
\end{figure}

\bluemark{Ca$_3$Ru$_2$O$_7$ crystallizes in the non-centrosymmetric space group \textit{Bb2$_1$m} \cite{kikugawa2007ca3ru2o7}, with associated point group \textit{mm2}, which includes a twofold rotation axis along $z$ and a vertical mirror plane. The full space group also contains a 2$_1$ screw axis along $z$, which plays a role in the nonsymmorphic symmetries relevant to the altermagnetic configuration. \cite{schiff2024collinear}.} The magnetic moment arises predominantly from the Ru 4$d$ electrons, in which, at low temperatures, the Ru atoms order ferromagnetically within the bilayers while aligning antiferromagnetically between them, forming an A-type antiferromagnetic structure where the spins are oriented along the $b$-axis (see Fig.~\ref{fig0}). Given this A-type antiferromagnetic order, the primitive cell of CRO contains spin-up sites, with a translational symmetry operation that maps spin-up into spin-down sites (along the $z$-axis). Consequently, the ground state is not altermagnetic (see Fig.~\ref{fig0}) \cite{vsmejkal2022beyond}.\\

\begin{figure}
\includegraphics[width=0.45\textwidth]{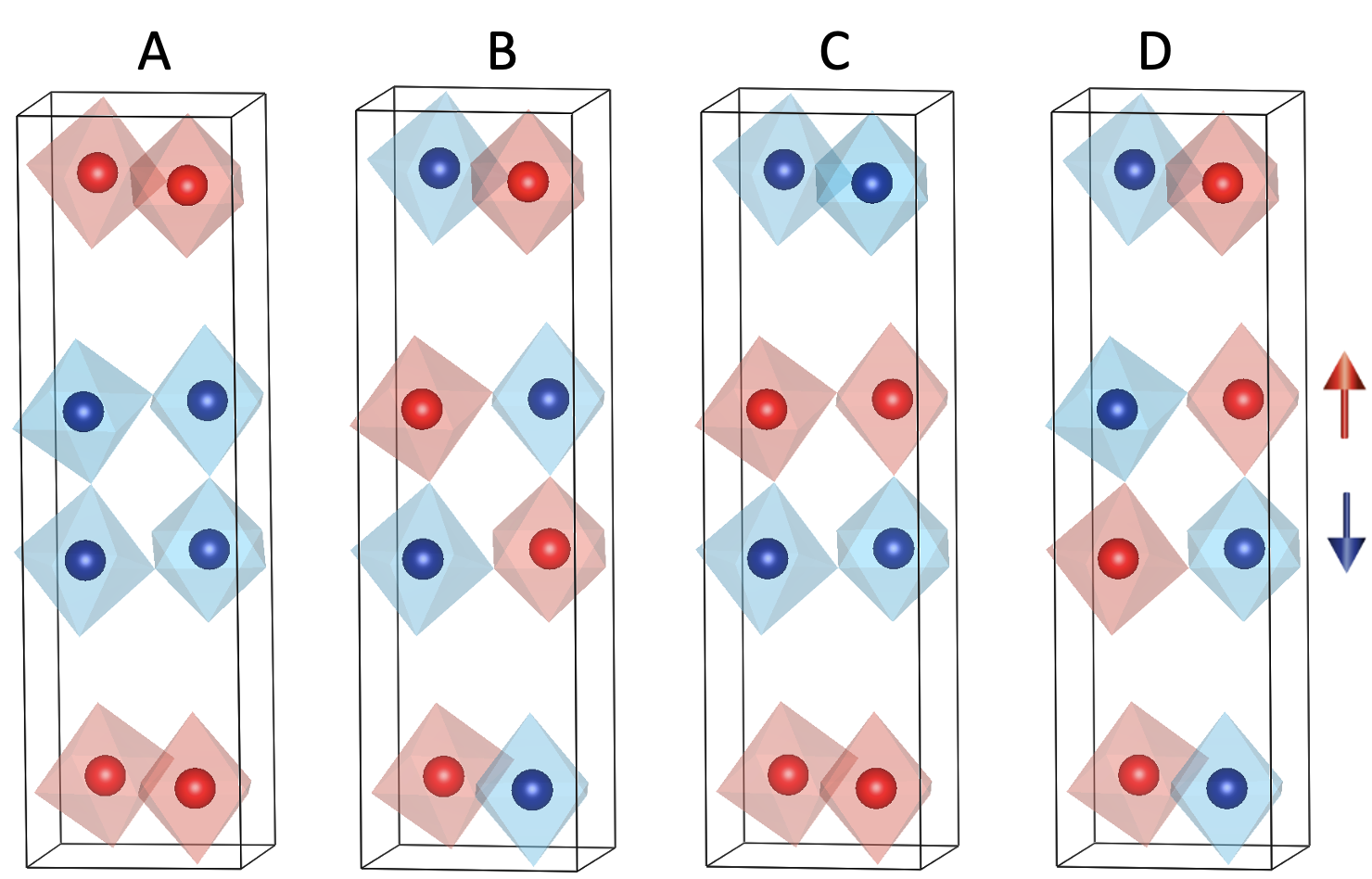}
\caption{Explored magnetic configurations in Ca\(_3\)Ru\(_2\)O\(_7\) with zero net magnetization in the non-relativistic limit. These are the only inequivalent magnetic phases. Red and blue spheres represent atoms with majority spin-up and spin-down orientation, respectively. Ca and O atoms are omitted for clarity.}
\label{fig:conf}
\end{figure}

We analyzed CRO under different magnetic configurations to explore possible AM states and their stability. Specifically, we examined magnetic orderings commonly observed in AB$_{3}$ perovskites \cite{gallego2012magnetic}, labeled as configurations A, B, C, and D (see Fig. \ref{fig:conf}). Table \ref{table:1} shows the energy differences relative to the ground-state configuration A, as well as some relevant structural, electronic, and magnetic properties for our study. Our results indicate that AM states appear in configurations B and C, where configuration B has an antiferromagnetic spin arrangement along the $a$-$b$ and $c$-axes, while configuration C shows this along the $c$-axis. Although these magnetic states have not been experimentally observed in CRO, similar phases have been reported in doped Ca$_{3}$Ru$_{2}$O$_{7}$ systems, such as Ti-doped Ca$_{3}$(Ru$_{1-x}$Ti$_{x}$)$_{2}$O$_{7}$ (configuration B) \cite{ke2011emergent}, and in the sister compound Sr$_{3}$Ru$_{2}$O$_{7}$ (configuration D) \cite{rivero2017predicting}.\\

Notably, variations in magnetic spin arrangements lead to significant structural and electronic changes (further structural details are given in SM table \ref{table:S1}). Specifically, FM and AFM in-plane arrangements result in metallic (A, C) and a narrow insular state (B, D), respectively. Next, we will focus on the AM phases of CRO, primarily in Configuration B.
\begin{table}[h!]
    \begin{tabular}{c|c|c|c|c}
        \hline
        \textbf{} & \boldmath{$\Delta$E} &\textbf{E-State}  &\textbf{M-Phase} & \boldmath{$m$ ($\mu_{B}$)} \\
        \hline
        A & 0   &  Metallic &AFM &  1.44\\
        \hline
        B & 34.90 & N-Insulator & AM  & 1.35  \\
        \hline
        C & 36.46  & Metallic & AM  & 1.38/1.45 \\
        \hline
        D & 39.22  & N-Insulator & AFM & 1.35\\
        \hline
    \end{tabular}
    \caption{The first column represents the label for each magnetic configuration. \textbf{$\Delta$E} is the energy difference relative to configuration A (E${_\text{conf.}}$ - E$_{A}$) (meV/Ru) without spin-orbit coupling; \textbf{E-phase} and \textbf{M-phase} indicate the electronic state and the AFM/AM ordering, respectively; and \textbf{\textit{m}} denotes the magnetic moment per Ru atom. Conf. A-C are metallic and Conf. B-D displays a Narrow (N) insulator state.}
    \label{table:1}
\end{table}

In addition to the electronic differences, configuration B exhibits a lower magnetic moment than configuration A (details on Table~\ref{table:1}). 
\bluemark{In configuration B, the magnetic moments form a N\'eel-type collinear ordering with alternating spin orientations. This promotes spin-compensated orbital overlap and wave function delocalization, which are characteristic of bonding-like states.} As a result, configuration B is more sensitive to the Hubbard-$U$ term due to the enhanced electronic proximity.
As a result, a band gap opens at relatively low U values, specifically $U = 1.0 \,\text{eV}$. In contrast, configuration A remains metallic under the same conditions. The Supplemental Material (Section \ref{DOS-U}) provides further analysis.
These contrasting results demonstrate CRO's highly correlated electronic nature, driven by the tight interplay between magnetic and structural degrees of freedom \cite{leon20243,gonzalez2025altermagnetism,puggioni2020cooperative,liu2011mott}. Finally, we confirm that these characteristics are predominantly electronic, arising from the magnetic spin configuration, and are independent of volume changes or structural distortions.\\
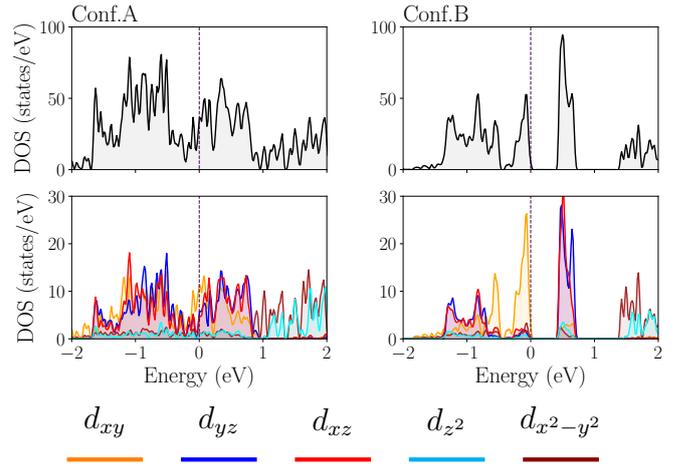
\begin{figure}[ht]
\includegraphics[width=0.245\textwidth]{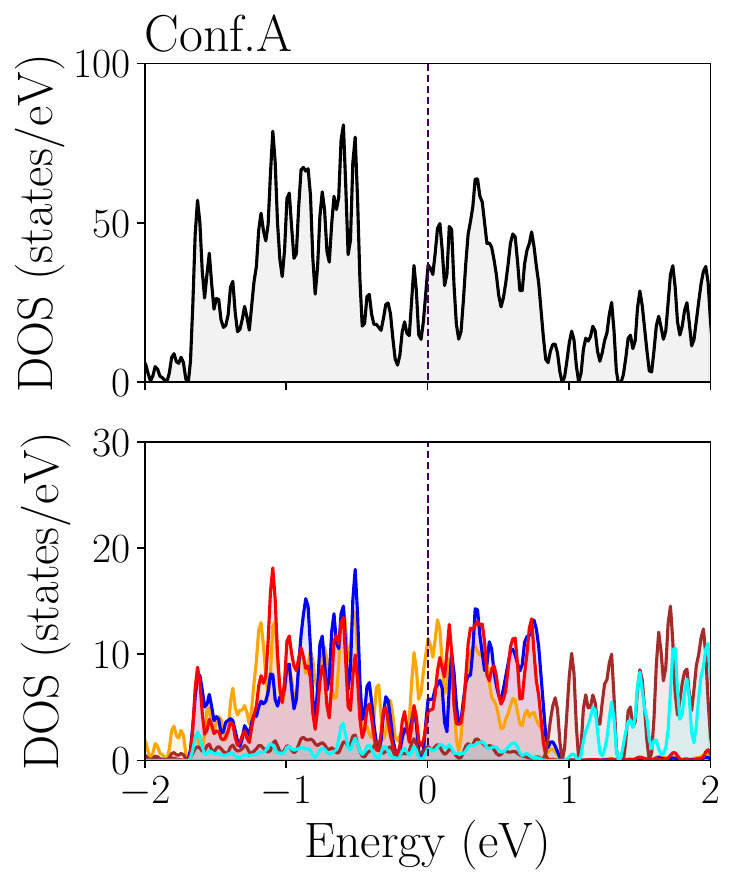}\includegraphics[width=0.245\textwidth]{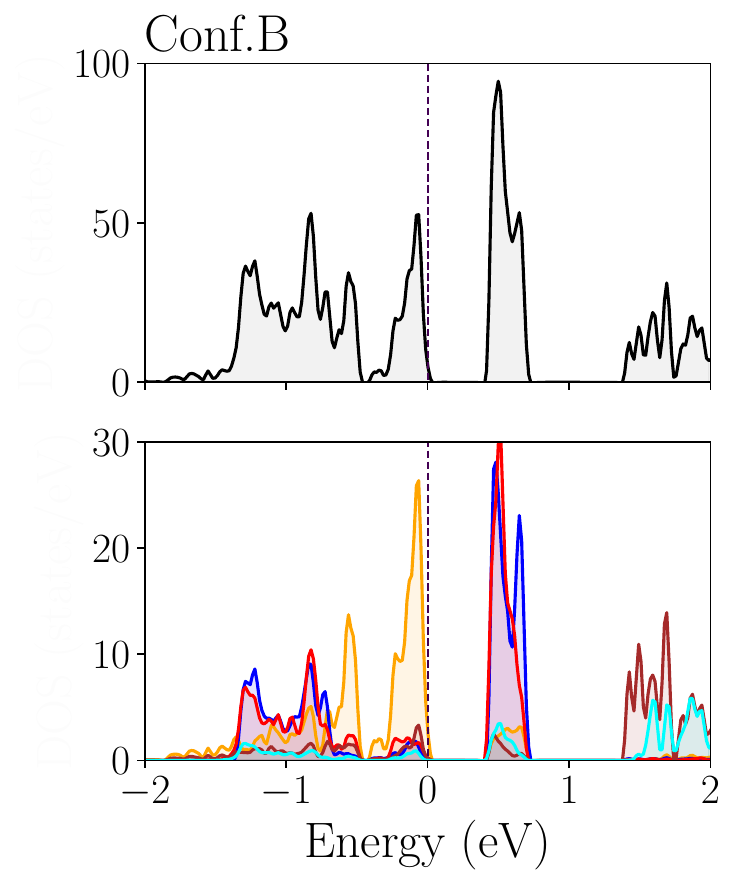}
\centering
 \begin{tikzpicture}
    \node at (0.5, 0.5) {\large $d_{xy}$};
    \draw[orange, ultra thick] (0,0) -- (1,0);
    
    \node at (2, 0.5) {\large $d_{yz}$};
    \draw[blue, ultra thick] (1.5,0) -- (2.5,0);
    
    \node at (3.5, 0.5) {\large $d_{xz}$};
    \draw[red, ultra thick] (3,0) -- (4,0);
    
    \node at (5, 0.5) {\large $d_{z^2}$};
    \draw[cyan, ultra thick] (4.5,0) -- (5.5,0);
    
    \node at (6.5, 0.5) {\large $d_{x^2-y^2}$};
    \draw[darkred, ultra thick] (6,0) -- (7,0);
\end{tikzpicture}
\caption{Upper and lower panels show the total and Ru-4\( d \) projected density of states (DOS), respectively, for magnetic configurations A and B.}
\label{fig:dos}
\end{figure}
\bluemark{We now examine the electronic structure of both configurations. Figure~\ref{fig:dos} shows the total and orbital-resolved density of states. While configuration~A is metallic, configuration~B exhibits a narrow gap at the Fermi level. In an ideal octahedral environment, the Ru $d_{xy}$, $d_{xz}$, and $d_{yz}$ orbitals are degenerate. Orthorhombic distortions lift this degeneracy slightly, but it is the N\'eel-type magnetic order in the $ab$-plane (conf.~B) that drives orbital differentiation by enhancing $\pi$-type hybridization between Ru-$d_{xy}$ and O-$p_{x}/p_{y}$ orbitals. This leads to a strong bonding--antibonding splitting, pushing the $d_{xy}$ bonding states to lower energy and making them nearly filled. The resulting spectral redistribution lowers the $d_{xy}$ chemical potential relative to the less-affected, half-filled $d_{xz}$ and $d_{yz}$ orbitals. The gap thus reflects an orbital-selective mechanism driven by the interplay between magnetism and crystal-field effects, as proposed for Ca$_2$RuO$_4$~\cite{cuono2023orbital}.}

\begin{figure*}[ht] 
{\includegraphics[width=1.0\textwidth]{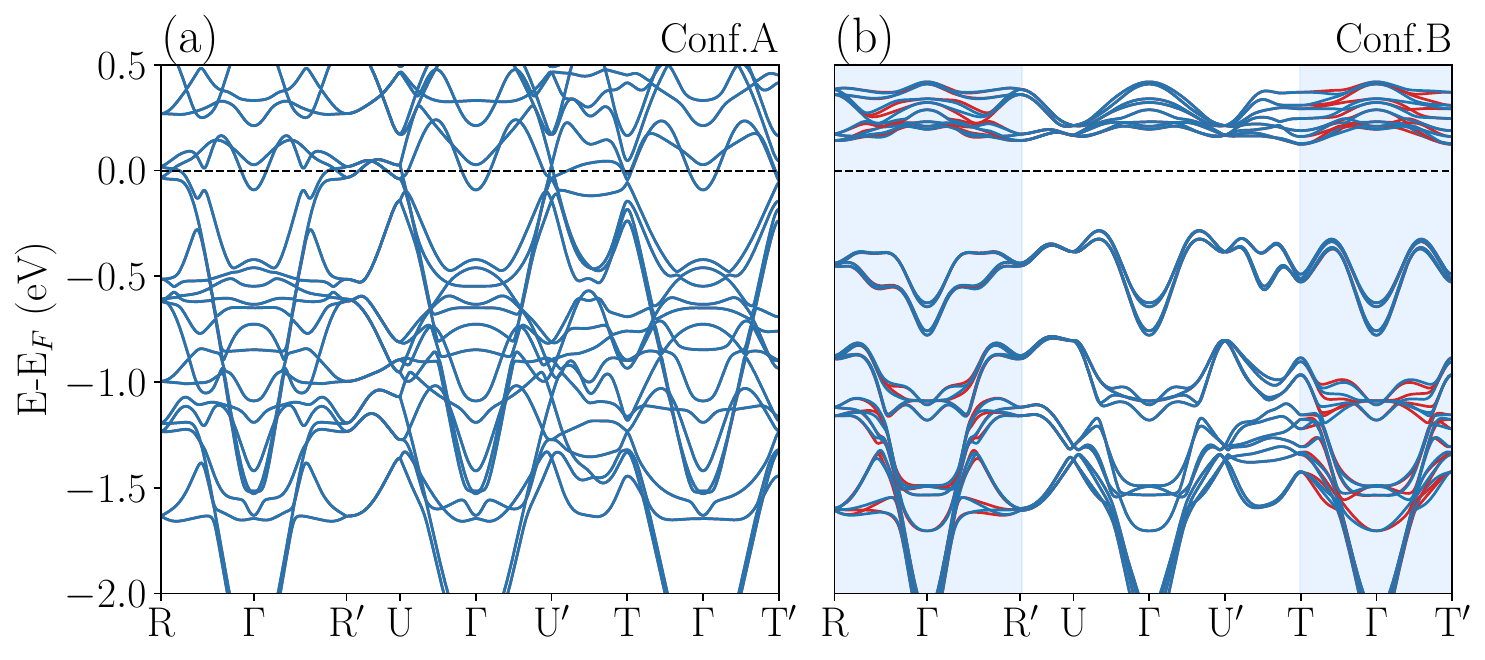}}
\centering
   \vspace{0.6cm} %
    \begin{tikzpicture}
        \node at (2, 0.5) {spin up};
        \draw[red, ultra thick] (1.5,0) -- (2.5,0);
        \node at (5, 0.5) {spin down};
        \draw[navy, ultra thick] (4.5,0) -- (5.5,0);
    \end{tikzpicture}
    \caption{(a)--(b) Electronic band structure for configurations A and B, respectively. The blue region highlights the altermagnetic bands. The AM region along the R--\(\Gamma\)--R\(^{\prime}\) and T--\(\Gamma\)--T\(^{\prime}\) directions. In contrast, the AFM region is defined along the \( yz \) direction, represented by the U--\(\Gamma\)--U\(^{\prime} \) path. The Brillouin zone is given in Fig. \ref{fig5}.}
    \label{fig:bandstructure}
\end{figure*}

\begin{figure}[ht] 
\centering
{\includegraphics[width=0.25\textwidth]{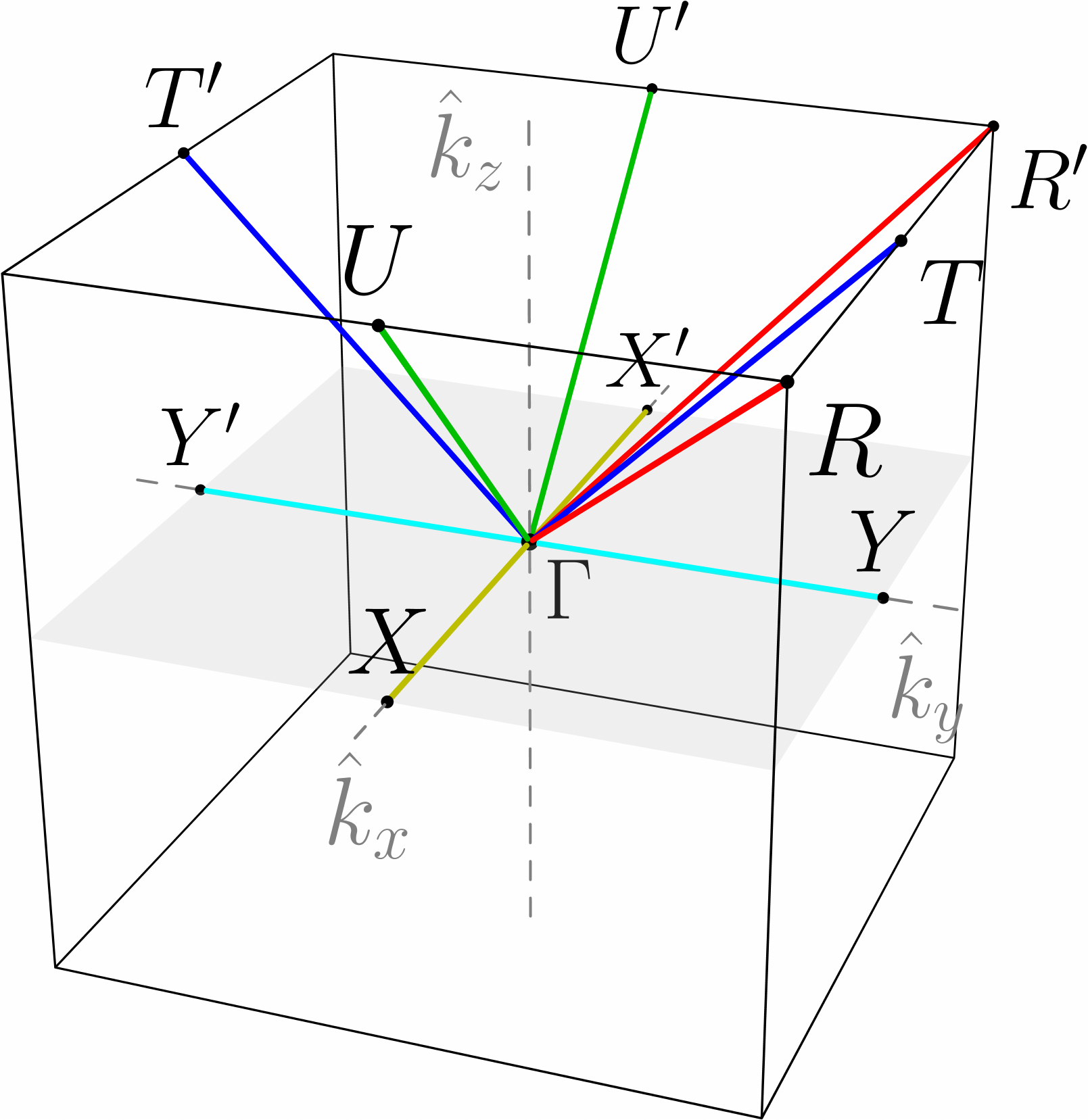}}
\caption{Brillouin zone of the orthorhombic structure. High-symmetry points related by symmetry are labeled as 
\( \mathrm{R}/\mathrm{R}' = (\pm 0.5, 0.5, 0.5) \), 
\( \mathrm{T}/\mathrm{T}' = (0,\pm 0.5, 0.5) \), 
\( \mathrm{U}/\mathrm{U}' = (\pm 0.5, 0,0.5) \), 
\( \mathrm{Y}/\mathrm{Y}' = (0, \pm 0.5, 0) \), and 
\( \mathrm{X}/\mathrm{X}' = (\pm 0.5, 0, 0) \), 
in units of the reciprocal lattice vectors.}
\label{fig5}
\end{figure}

Figure~\ref{fig:bandstructure} shows the non-relativistic band structure for the A and B magnetic configurations. The A configuration (ground state) exhibits metallic behavior, consistent with previously reported results that neglect SOC effects \cite{markovic2020electronically,leon20243}. In contrast, configuration B displays a narrow gap and a $k$-dependent non-relativistic spin splitting, a hallmark of altermagnetic states (see Fig. \ref{fig:bandstructure}(b)). The shaded regions on the band structure indicate regions in $k$-space where the altermagnetic spin splitting is maximal. Figure~\ref{fig5} shows the detailed Brillouin zone (BZ) corresponding to these paths.

\bluemark{The presence of the non-relativistic spin-splitting is independent of the value of k$_y$. The observed spin splitting exhibits what is known as a d-wave symmetry in momentum space. This terminology refers to how the splitting changes sign across the Brillouin zone, in analogy to the lobes of a d-orbital. Specifically, the splitting changes sign upon inversion of the k$_x$ or k$_z$ coordinate, but remains unchanged when inverting k$_y$. This independence from the k$_y$ component allows for a more precise classification as a d$_{xz}$-wave character. According to the formal symmetry classification of altermagnets \cite{vsmejkal2022beyond}, this behavior corresponds to a P-2 d$_{xz}$-wave state. These paths, which extend beyond the xy plane, have not been explored in previous studies, where ARPES experiments primarily focused on the xy plane \cite{markovic2020electronically,horio2021electronic}. The following section discusses the preference for a specific direction in the AM character.}

\begin{figure*}[ht]
\includegraphics[height=4.4cm]{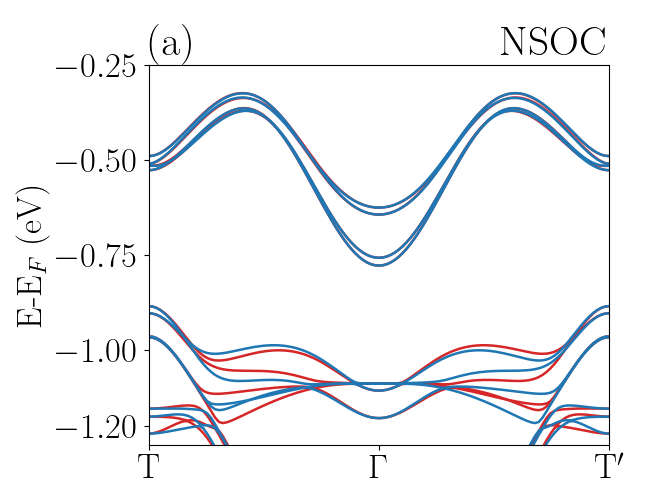}
\includegraphics[height=4.38cm,trim=1.5cm 0cm 2.5cm 0cm, clip]{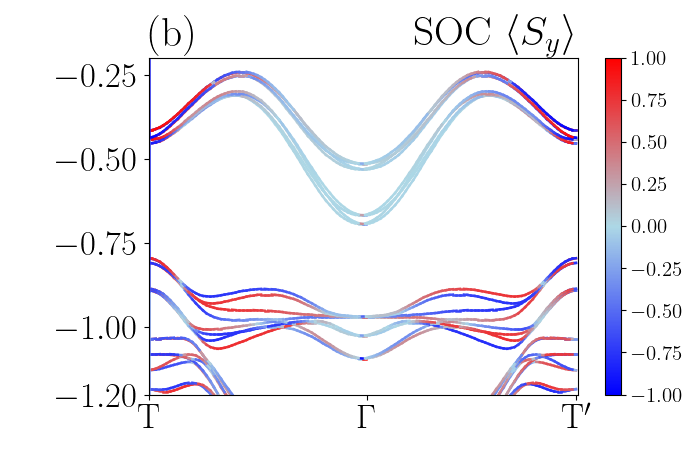}
\includegraphics[height=4.38cm,trim=1.5cm 0cm 0cm 0cm, clip]{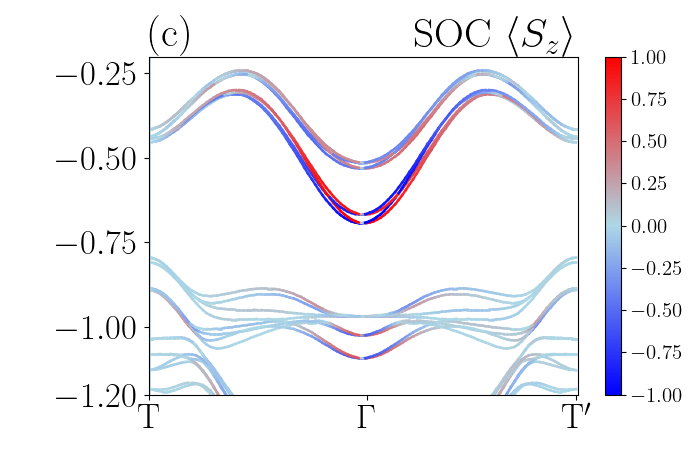}
\includegraphics[height=4.48cm]{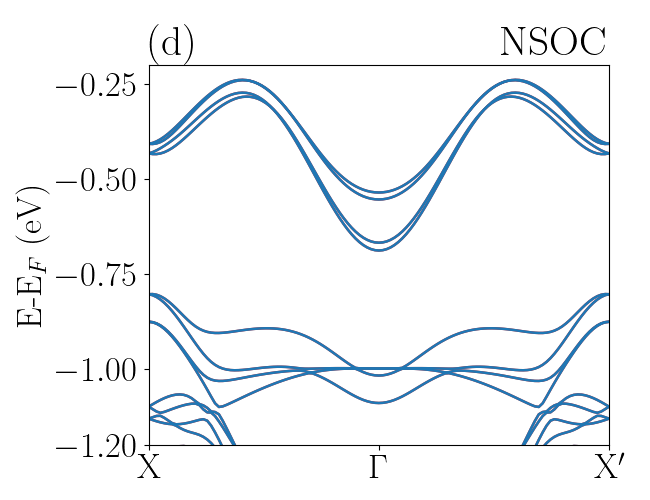}
\includegraphics[height=4.38cm,trim=1.0cm 0cm 2.5cm 0cm, clip]{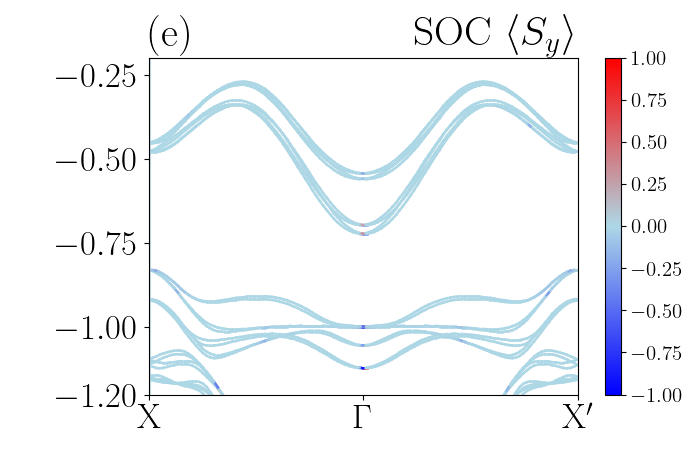}
\includegraphics[height=4.38cm,trim=0.6cm 0cm 0cm 0cm, clip]{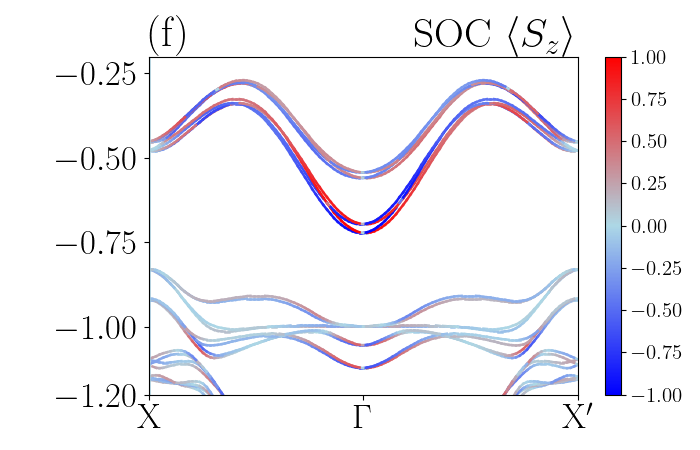}
\caption{Comparison of band structures for Ca$_{3}$Ru$_{2}$O$_{7}$ without (NSOC) and with (SOC) spin-orbit coupling. 
Top row: Band structure along the \(\mathrm{T}-\Gamma-\mathrm{T}'\) path. 
Bottom row: Band structure along the \(\mathrm{X}-\Gamma-\mathrm{X}'\) path. 
For both paths, panels show the NSOC case (a,d) and SOC projections along \(S_y\) (b, e) and \(S_z\) (c, f). The color scale indicates the spin expectation value (red: positive, blue: negative). The full energy range is shown in Fig.~\ref{fig:S4}.}
\label{fig_soc}
\end{figure*}

The bands located between \([-2.0, -1.0]\)~eV exhibit significant non-relativistic spin splitting, corresponding to the \(d_{xz}\)-\(d_{yz}\) orbitals. In contrast, the bands between \([-0.7, -0.5]\)~eV, dominated by the \(d_{xy}\) orbital character, show negligible spin splitting (see, Fig. \ref{fig:S2}). Therefore, configuration B exhibits orbital-selective altermagnetism with the altermagnetism present in the \(d_{xz}\)-\(d_{yz}\) bands but not in the \(d_{xy}\) bands, this is similar to that observed in Ca\(_2\)RuO\(_4\)~\cite{cuono2023orbital}. When the \(k_x\) or \(k_z\) coordinate is inverted in reciprocal space, the non-relativistic spin splitting changes sign, while inversion of \(k_y\) leaves it unchanged. Notably, this altermagnetic order matches that found in Ca\(_2\)RuO\(_4\), despite its crystallization in a different space group~\cite{cuono2023orbital}. The orbital-selective magnetism depends on the magnetic space group and orbital character. In conf.B, the nearly filled d$_{xy}$ orbitals do not contribute to magnetism, leading to orbital-selective AM. Consequently, conf. C displays different magnetic states in which all the t$_{2g}$ orbitals contribute to the AM character; therefore, it does not exhibit orbital selective AM, see more details in Fig. \ref{fig:S2}.\\

The lack of inversion symmetry in Ca$_3$Ru$_2$O$_7$ enables sizable SOC effects, which play a central role in its magnetic properties \cite{markovic2020electronically}. In configuration B, as in configuration A, the spins align primarily along the $y$-axis, with a slight canting toward $z$-axis ($(0,\,1.3,\,0.1)\,\mu_\mathrm{B}$). \bluemark{ SOC enriches the spin textures and activates symmetry-breaking mechanisms that lift band degeneracies, thereby enabling access to AM states that remain hidden in the collinear case.}

\bluemark{Our spin-projected band structure analysis reveals how SOC qualitatively transforms the nature of altermagnetism in this system. The non-relativistic ground state exhibits a clear $d$-wave character, dominated by $d_{xz}$-like states along the high-symmetry paths shown in Fig.~\ref{fig:bandstructure}. Upon including SOC, a complex landscape emerges. The original $d$-wave features remain most evident in the $\langle S_y \rangle$ projection (Figs.~\ref{fig_soc}(a)-(b)), consistent with the magnetic anisotropy along the $b$-axis. In stark contrast, the $\langle S_z \rangle$ projection unveils a completely new altermagnetic signature between -0.5 and -0.75\,eV, emerging from bands that were purely antiferromagnetic in the collinear case (Figs.~\ref{fig_soc}(a),(c)).
The influence of SOC is particularly evident along the $X-\Gamma-X'$ direction. Here, SOC lifts the band degeneracy in the $\langle S_z \rangle$ channel, creating a spin splitting where the non-relativistic bands were degenerate (Figs.~\ref{fig_soc}(e)-(f)) (similar results are found for \( \langle S_x \rangle \) along \( k_z \)). This purely relativistic splitting exhibits a characteristic antisymmetric behavior for momentum inversion ($\mathbf{k} \rightarrow -\mathbf{k}$), which is the defining hallmark of a $p$-wave altermagnetic component.}

\bluemark{Therefore, these findings demonstrate the emergence of a novel hybrid $d/p$-wave altermagnetic order. The AM+SOC phase in CRO is best described as a state where the non-relativistic $d$-wave component coexists with a purely SOC-driven $p$-wave character. This hybridization arises from the interplay between the magnetic structure and the crystal's nonsymmorphic symmetries.}

While these results highlight a recently emerging regime of altermagnetism in the presence of SOC \cite{cheong2024altermagnetism,vsmejkal2411altermagnetic}, a formal classification of this hybrid order, for instance, through advanced group-theoretical analysis, remains a substantial theoretical challenge. we describe the emergent AM+SOC features through their direct, symmetry-resolved signatures in the spin-projected electronic structure. Future studies will address a more systematic classification of AM+SOC spin textures.

\subsection*{Symmetry analysis}
\begin{figure*}[ht]
    \centering
    \raisebox{4mm}{\includegraphics[width=0.32\textwidth]{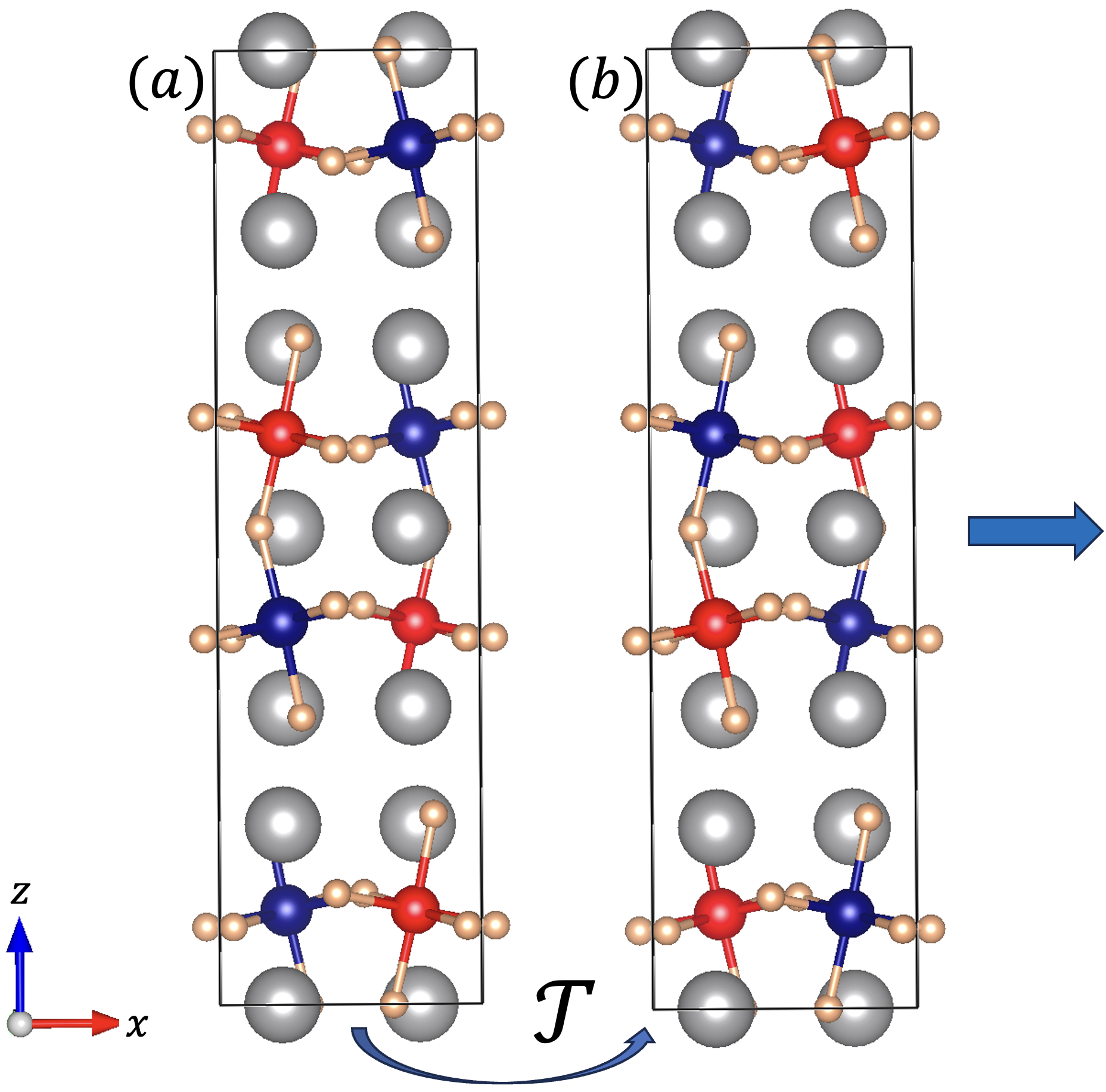}} \includegraphics[width=0.58\textwidth]{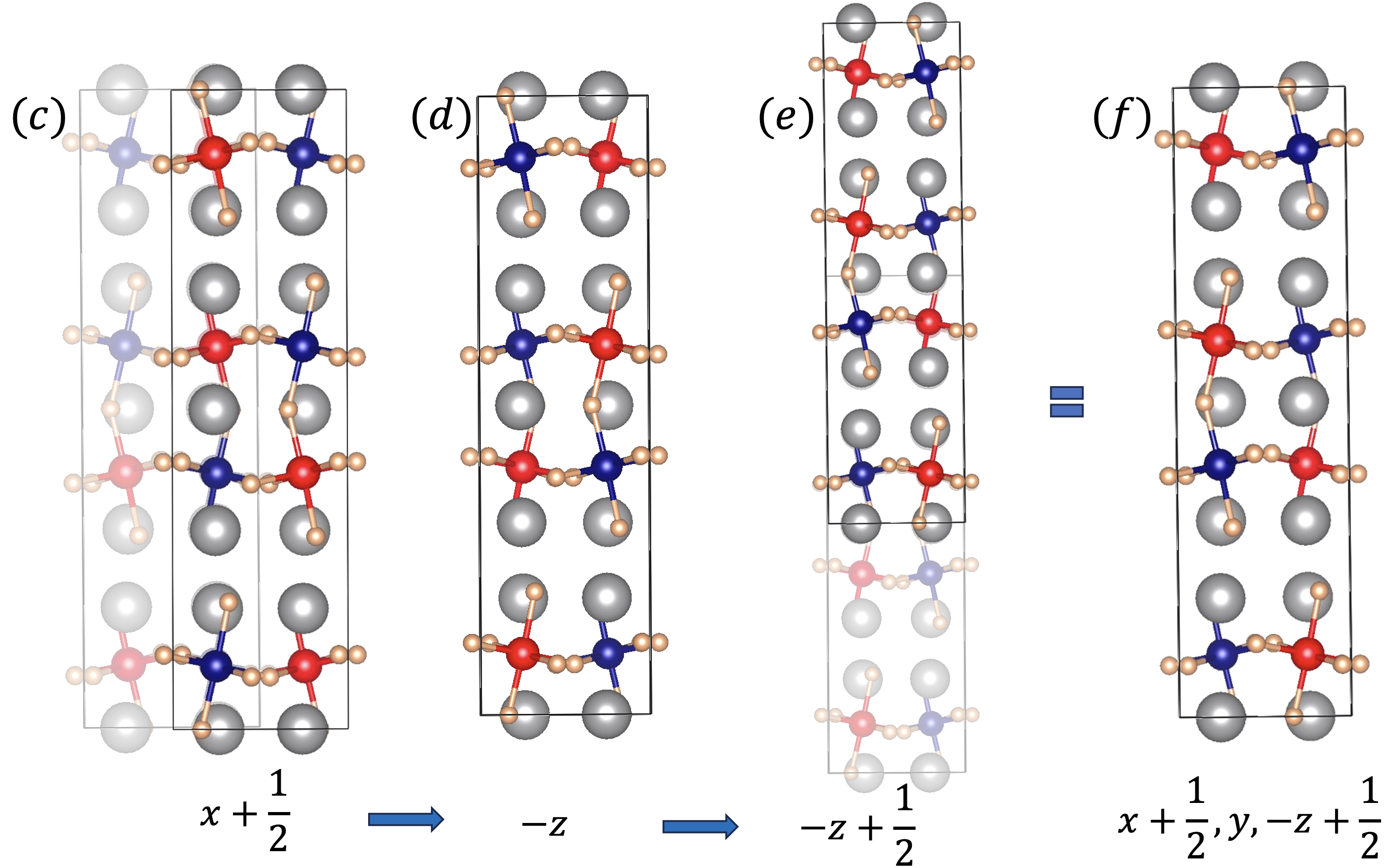}
    \caption{Symmetry operations responsible for the altermagnetic character of Ca\(_3\)Ru\(_2\)O\(_7\) in the B magnetic configuration. \(\mathcal{T}\) denotes the time-reversal operator, which interchanges spin-up and spin-down states, represented by red and blue spheres, respectively.}
    \label{fig:symmetry}
\end{figure*}
To understand the role of symmetry in AM-CRO, consider applying a time-reversal operator (TR), which in practical terms consists of an exchange between spin-up and spin-down (see Fig.~\ref{fig:symmetry}(a)-(b). Due to the distinct symmetry properties of many altermagnets, the original magnetic configuration cannot be recovered using only rotational symmetry operations. Following a time-reversal operation, a half-unit cell translation along the $x$-axis, shifting $x \rightarrow x + 0.5$ (Fig.~\ref{fig:symmetry}(c)), is combined with a screw symmetry operation by inverting $z \rightarrow -z$ and adding a half-unit cell translation along the $z$-axis (Fig.~\ref{fig:symmetry}(d)-(e)). 
In summary, the operation $(x + \frac{1}{2}, y, -$z$ + \frac{1}{2})$ becomes necessary (Fig.~\ref{fig:symmetry}(f)). These combined nonsymmorphic operations result in the $k$-dependent splitting states \cite{vsmejkal2022beyond}. Due to the symmetry operations that connect the spins, which are determined by rotations and translations along the $x$ and $z$ directions while leaving the $y$ direction unchanged, the AM character is found along these axes, as shown in Fig.~\ref{fig:bandstructure}. In the case of the C configuration, the symmetry operation ($x$ +$\frac{1}{2}$, $y$ + $\frac{1}{2}$, -$z$ +  $\frac{1}{2}$) becomes necessary to achieve AM order. \bluemark{This symmetry operation is specific to the orthorhombic space group Bb2$_{1}$m. Since the crystallographic axes are not symmetry-equivalent, this operation does not remain valid under the replacement \( x \rightarrow y \)
}.\\

\subsection*{Tuning AFM to AM phases}

\begin{figure}[ht]
\includegraphics[width=0.45\textwidth]{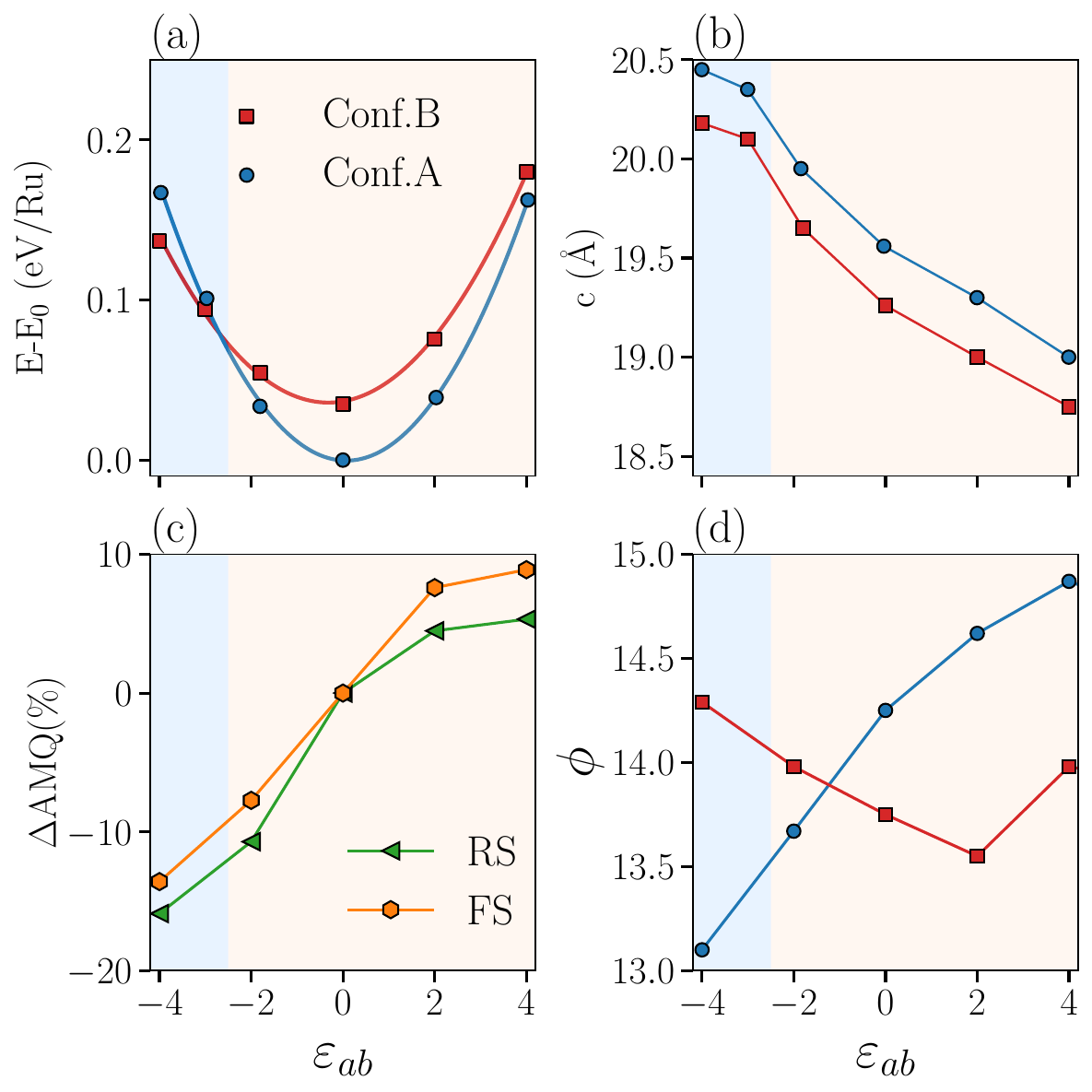}
\centering \includegraphics[width=0.31\textwidth]{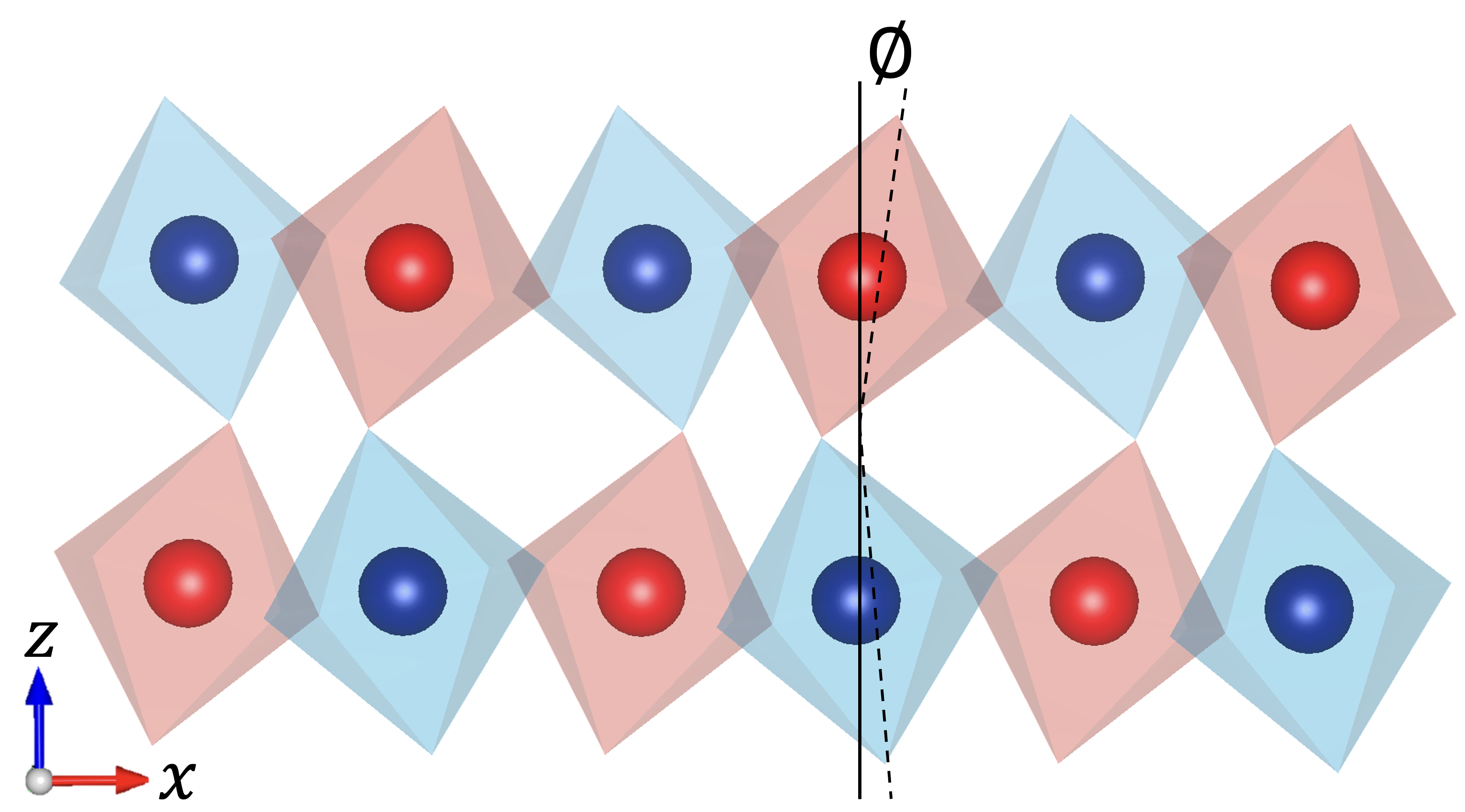}
\caption{
(a) Total energy variation \( (E - E_0^A) \) as a function of biaxial strain \( (\varepsilon_{ab}) \) for configurations A and B without spin--orbit coupling.
(b) Evolution of the \( c \)-lattice parameter as a function of strain.
(c) Variation of the Altermagnetic Merit Quantity (AMQ), relative to the unstrained system, as defined in Eq.~\ref{eq:3}. Hexagons and triangles represent calculations performed under relaxed-structure (\textbf{RS}) and fixed-structure (\textbf{FS}) conditions, respectively, for configuration B.
(d) Evolution of the RuO octahedral distortion angle \( (\phi) \) for configurations A and B.
Blue (red) shaded regions indicate strain ranges where the AM (AFM) phases are energetically favored.
}
\label{fig6}
\end{figure}

To analyze the stability of configurations A and B, we calculated their total energies under various biaxial strain conditions (Fig.~\ref{fig6}(a)). Configuration B shows the lowest energy within a compressive strain below $-2\%$. Fig.~\ref{fig6}(b) shows the $c$-lattice parameter for each strain. These results suggest a potential magnetic transition from AFM to AM states in CRO.

\begin{figure*}[h]
\centering
 \begin{tikzpicture}
    \node at (2.5, 0.5) {spin up};
    \draw[red, ultra thick] (2,0) -- (3,0);
    \node at (5, 0.5) {spin down};
    \draw[blue, ultra thick] (4.5,0) -- (5.5,0);
\end{tikzpicture}\\
\vspace{0.2cm}
\includegraphics[width=0.96\textwidth]{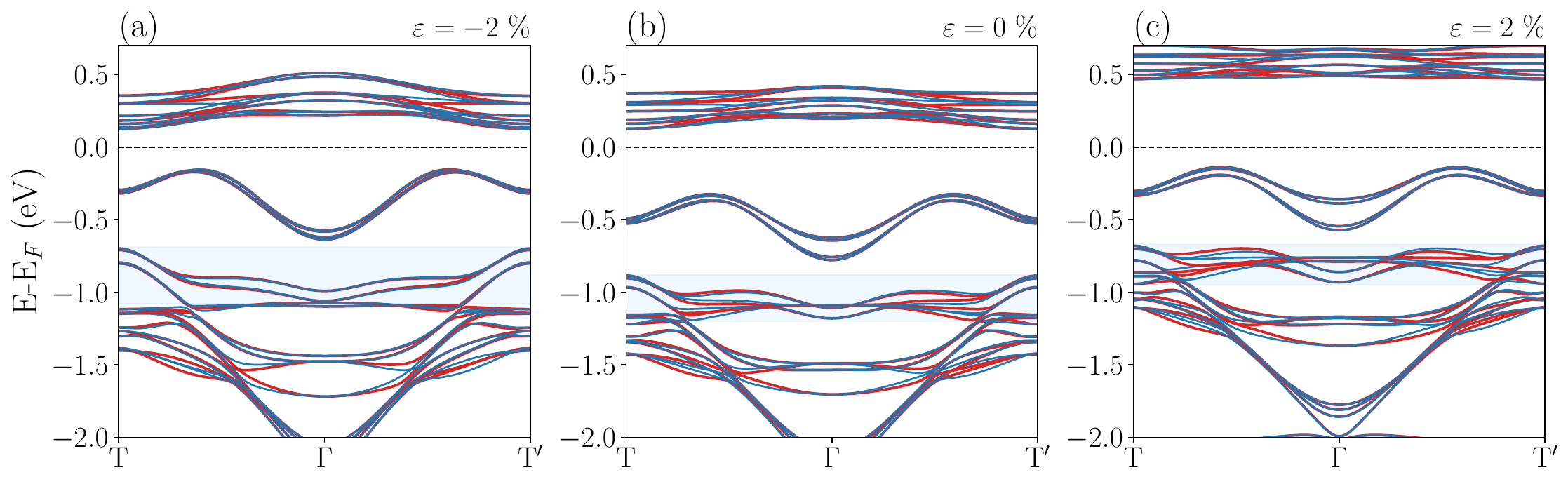}
\includegraphics[width=0.96\textwidth]{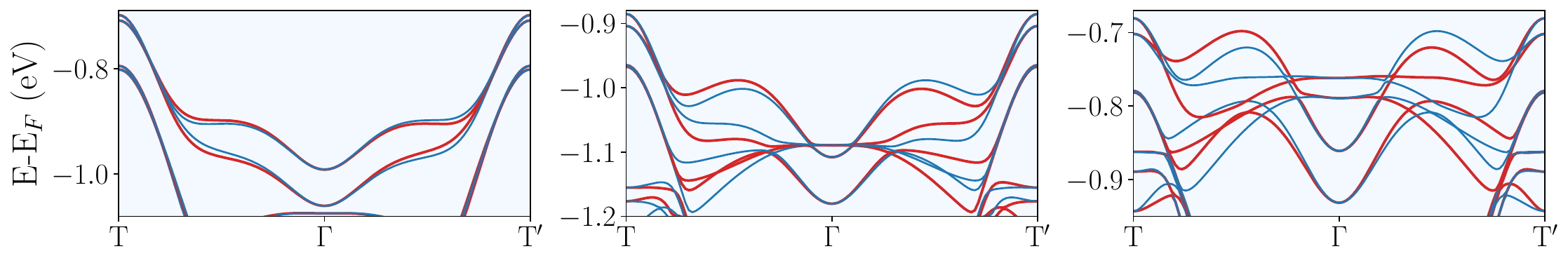}
\caption{Top row: Non-relativistic band structures of configuration B under three strain conditions: (a) $-2\%$ compressive, (b) $0\%$ (unstrained), and (c) $+2\%$ tensile, plotted along the selected $\mathrm{T}-\Gamma-\mathrm{T}'$ path. Bottom row: zoomed-in view of the region showing the strongest altermagnetic features, which varies with strain.}
\label{fig7} 
\end{figure*}

Next, we investigate  the effects of compressive (-$\varepsilon$) and tensile (+$\varepsilon$) strain on the AM features of CRO. Fig.~\ref{fig7} shows the band structure of configuration B under strains of $\varepsilon$ = -2\%, 0, and 2\% (biaxial). It can be observed that compressive strain suppresses the AM features among the energy range [-2:-1] eV (compare Fig.~\ref{fig7} (a) with Fig. \ref{fig7}(b)). Moreover, the tensile strain enhances the splitting and shifts it to higher energies.\\

To quantify the effect of strain on the altermagnetic features of CRO, we calculate the average difference between spin-up and spin-down eigenvalues \(\Delta E(\mathbf{k})\) for each occupied band along the entire \(\mathbf{k}\)-path. Additionally, we define the altermagnetic quantity (AMQ) as the altermagnetic figure of merit, which integrates \(\Delta E(\mathbf{k})\) over the Brillouin zone. The AMQ is a numerical measure of the overall altermagnetic character of a system, providing a quantitative basis for comparing different strain conditions. 

In our implementation, we read the spin-resolved eigenvalues from the \texttt{EIGENVAL} file generated by VASP, then compute the \(\mathbf{k}\)-dependent average splitting \(\langle \Delta E(\mathbf{k})\rangle\) 
as the cumulative sum over occupied states along a particular set of k points (over a \(\mathbf{k}\)-path or \(\mathbf{k}\)-grid), given by:
\begin{equation}
\langle \Delta E(\mathbf{k})\rangle \;=\; \sum_{j=1}^{N_{occ}}\!
\left|
 \frac{1}{N_k}\sum_{i=1}^{N_k} 
   \bigl[E_{\text{down}}(\mathbf{k}_{i,j}) \;-\; E_{\text{up}}(\mathbf{k}_{i,j})\bigr]
\right| \,,
\label{eq1}
\end{equation}
where N$_{k}$ is the numbers of bands at $k$-point within specified energy range and N$_{occ}$ is the occupied bands.\\

Note that the integration to obtain the AMQ can be performed either over the entire Brillouin zone or along a specific \(\mathbf{k}\)-path. We verified that both methods yield similar solutions; the only difference is a constant multiplicative factor, which cancels out when computing percentage changes.

We sum up all the occupied states rather than comparing only a few selected bands. This approach proves more reliable because the most strongly split bands can vary across different systems and even within the same material under strain, depending on details such as band character and crystal symmetry. Consequently, integrating over the entire set of occupied bands offers a more robust assessment of the altermagnetic spin-splitting.

The AMQ quantifies the spin-splitting across momentum space by integrating \(\langle \Delta E(\mathbf{k})\rangle\):
\begin{equation}
\text{AMQ} = 
\int_{\text{BZ}} \langle \Delta E(\mathbf{k}) \rangle \, d\mathbf{k}\,,
\label{eq:2}
\end{equation}
in our code, Simpson’s rule is used for numerical integration. When \(\mathbf{k}\) points are expressed in inverse angstroms (\(\text{\AA}^{-1}\)), the resulting AMQ is in units of \(\mathrm{eV}\cdot\text{\AA}\). 
\bluemark{Although the AMQ is not a formal symmetry-derived order parameter like magnetic multipoles \cite{bhowal2024ferroically}, it provides a symmetry-consistent and practical metric to quantify momentum-resolved spin splitting in systems with zero net magnetization. While this approach is robust, it may fail near band crossings or overlapping bands. A more rigorous treatment, such as disentanglement via Wannier-based orbital tracking, could refine the analysis but lies beyond the scope of this study. Nonetheless, the observed vanishing of spin splitting at high-symmetry points and the overall internal consistency strongly support the identification of altermagnetic behavior.}

Since our primary goal is to evaluate strain effects on the altermagnetic state, we compare the AMQ under strain to the unstrained scenario. We define the percent variation \(\Delta \text{AMQ}\) as:
\begin{equation}
\Delta \text{AMQ} \;=\; 
\frac{\text{AMQ}(\varepsilon) - \text{AMQ}(\varepsilon=0)}{\text{AMQ}(\varepsilon=0)} 
\times 100 \,.
\label{eq:3}
\end{equation}
This relative measure properly captures how strain modifies the degree of altermagnetic spin-splitting throughout the Brillouin zone.\\

Fig.~\ref{fig:S6} shows the evolution of the spin splitting \( \langle \Delta E(k) \rangle \) across the full \( k \)-path and occupied bands under different strain conditions (see Brillouin zone in Fig.~\ref{fig:bandstructure}). The magnitude of band-dependent altermagnetic splitting varies with strain, increasing notably under tensile strain. The AQM response varies by approximately $+8\%$ under tensile strain (2\%) and $-11\%$ under compressive strain (-2\%), as shown in Fig.~\ref{fig6}(c). All computational details and equations are provided in the Supplementary Information (Fig.~\ref{fig:S7}).\\

To further dissect the symmetry of the spin splitting in three dimensions (details provided in the SM, section SVI), we perform a multipolar analysis of the energy texture $\Delta E^{\beta}(\mathbf{k})$. In the non-relativistic limit, the response is purely even in $\mathbf{k}$, confirming the expected pure $d$-wave altermagnetic state dictated by symmetry. Including SOC activates odd-parity components and transforms the system into a hybrid state with concurrent $p$-, $d$-, and $f$-wave character. Using spinor band pairs near the Fermi level, we find overall $p$-wave dominance, with fractional $L^1$-norm weights $f_p \approx 0.47$, $f_d \approx 0.29$, and $f_f \approx 0.24$. The hybridization is strongly anisotropic across spin channels: the $x$ and $z$ components are $p$-dominated ($f_p^x \approx 0.59$, $f_p^z \approx 0.60$), whereas the $y$ component is $d$-dominated ($f_d^y \approx 0.59$).

\section*{Discussion}

The non-relativistic spin-splitting in CRO arises from the interplay between  hopping parameters and the opposite on-site spin polarizations in the two magnetic sublattices. Strain introduces competing effects that influence these parameters: compression along one axis promotes electronic delocalization, weakening magnetism and reducing the AM band splitting, whereas tensile strain favors localization, enhancing magnetic exchange and increasing the AM band splitting.

From the energy-strain relationship shown in Fig.~\ref{fig6}(a), we observe that configuration B (the AM phase) becomes energetically more favorable than configuration A (the AFM phase) under biaxial compressive strain (\(\varepsilon_{ab} < -2\%\)). Compressive strain decreases the in-plane Ru-O bond lengths, forcing an expansion along the \(c\)-axis. This expansion modifies the balance of structural distortions and stabilizes configuration B over configuration A. At the electronic level, the in-plane compression broadens the bands, weakening the magnetic exchange interactions and consequently reducing the AM band splitting. Conversely, under tensile strain, the lattice is stretched in the \(ab\)-plane while slightly contracting along \(c\), effectively narrowing the bandwidth and reinforcing the altermagnetic character (Fig. \ref{fig:S5} SM). This effect is reflected in the rising AMQ values in Fig.~\ref{fig6}(c) RS line (relaxed structures), demonstrating how strain can selectively enhance or suppress altermagnetic character.\\

To further assess the role of structural distortions, we performed an additional set of calculations where the atomic positions remained fixed at their zero-strain configuration. At the same time, only the lattice parameters were modified (fixed structure (FS) calculations). The results, depicted by the FS line in Fig.~\ref{fig6}(c), reveal a similar overall trend: AMQ increases under tensile strain and decreases under compressive strain. However, for \(\varepsilon_{ab} > 2\%\), the deviation between the relaxed and non-relaxed calculations becomes more pronounced, suggesting that octahedral distortions (\(\phi\)) play a secondary but non-negligible role in the strain response of the AM phase. Specifically, as seen in Fig.~\ref{fig6}(d), \(\phi\) increases under compressive strain for configuration B, further enhancing localization effects that help sustain the AM band splitting.
A direct comparison between the relaxed structures and the non-relaxed structures allows us to disentangle the effects of strain-induced delocalization/localization from those arising purely from octahedral distortions. In particular, the density of states in Fig.~\ref{fig:S5} confirms that compressive strain broader states, reducing AM features. In contrast, tensile strain narrows the bandwidth, reinforcing the AM character. Moreover, the band structures shown in Fig.~\ref{fig:S7} illustrate how freezing the atomic positions primarily alters the gap structure rather than the AM spin splitting itself, reinforcing the observation that the observed strain-dependent AM behavior is primarily dictated by electronic exchange interactions, with octahedral distortions acting as a secondary fine-tuning mechanism.\\

Furthermore, Fig.~\ref{fig6}(d) reveals how the RuO octahedral distortion angle $\phi$ responds to biaxial strain in each magnetic configuration. In configuration A, $\phi$ increases smoothly from $-4\%$ to $+4\%$ strain. Configuration B shows a more intricate pattern: under compressive strain ($\varepsilon_{ab}<0$), $\phi$ increases as the lattice compresses in the plane. This response reflects stronger electronic localization in the antiferromagnetic arrangement (see Fig.~\ref{fig:dos}(b)), which preserves magnetic exchange by enhancing octahedral distortion. Under tensile strain, in-plane expansion tends to relax distortions while the slight contraction along the $c$ axis reinforces them. The net change in $\phi$ depends on the balance between these two opposing effects.

The evolution of $\phi$ in configuration B correlates directly with the energy profile in Fig.~\ref{fig6}(a). Above $2\%$ compressive strain, configuration B becomes lower in energy than configuration A and $\phi$ increases sharply. This coincidence demonstrates that enhanced octahedral distortion drives the AFM-to-AM transition. \bluemark{Although such strain levels are challenging to achieve in bulk crystals, epitaxial oxide films can routinely accommodate comparable lattice mismatches, providing a practical platform to realize and probe distortion-driven magnetic and electronic phase transitions in layered perovskites}~\cite{wang2024transport,burganov2016strain,dong2024polar,tsubaki2023significant}.

This work investigates the emergence of altermagnetic states in the correlated oxide Ca$_3$Ru$_2$O$_7$. While its ground state is not altermagnetic, symmetry-allowed AM order arises in alternative spin configurations. In the non-relativistic case, Configuration~B, characterized by a N\'eel-type spin pattern, hosts a P-2 \(d_{xz}\)-wave altermagnetic state.

\bluemark{Including spin-orbit coupling with the N\'eel vector along $b$ induces additional symmetry breaking, allowing a $p$-wave component in the spin texture. This hybridization between $d$- and $p$-wave features may result from the absence of inversion symmetry and the presence of nonsymmorphic operations. A complete understanding of the $p$-wave origin will require further analysis based on spin space group symmetries under non-collinear magnetism.}

Although configuration~B has not been reported in pristine Ca$_3$Ru$_2$O$_7$, it appears in doped compounds such as Ca$_3$(Ti$_x$Ru$_{1-x}$)$_2$O$_7$, suggesting that suitable conditions can stabilize the AM state. Our results further demonstrate that biaxial strain acts as an effective tuning parameter for the altermagnetic band splitting: tensile strain (or equivalently, compression along the \( c \)-axis) enhances the splitting by approximately \( 9\% \) at \( 2\% \) strain, whereas compressive strain reduces it by about \( 11\% \). We also predict a strain-induced transition from the AFM ground state to the AM state beyond 2\% compressive strain. While prior studies have used uniaxial strain to manipulate magnetic anisotropy~\cite{dashwood2023strain}, the effects of biaxial strain remain largely unexplored.\\

Inducing altermagnetism in Ca$_3$Ru$_2$O$_7$ would create a rare noncentrosymmetric altermagnetic material and provide a platform to explore the interplay between odd- and even-wave spin-momentum locking. Identifying such phases would broaden the spectrum of quantum phenomena known in the Ca$_{n+1}$Ru$_n$O$_{3n+1}$ family, which already exhibits strain-tunable transitions, including Mott suppression in Ca$_2$RuO$_4$ ($\sim$ $-4\%$ strain)\cite{ricco2018situ}, the emergence of a Kondo effect, and a metal-to-semiconductor transition in CaRuO$_3$ ($-3.6\%$ strain)\cite{wang2024transport}. Our results highlight biaxial strain as a promising route to control altermagnetic phases in Ca$_{n+1}$Ru$_n$O$_{3n+1}$-related materials, aligning with recent proposals to manipulate antiferromagnetic-altermagnetic phase transitions via strain \cite{chakraborty2024strain,xun2025stacking}. These findings motivate further exploration of symmetry-driven spin textures in systems with strong spin-orbit coupling.\\

\textit{Note:} Recent studies have shown that SOC can induce interesting phenomena such as polarization switching in Ca$_3$Mn$_2$O$_7$, driven by magnetic anisotropy~\cite{vsmejkal2411altermagnetic}. Given its structural similarity to Ca$_3$Ru$_2$O$_7$, our results suggest that Ca$_3$Mn$_2$O$_7$ may also host mixed-symmetry altermagnetic order under SOC. Further exploration of this system could offer new pathways for tuning altermagnetic behavior via SOC.

\section*{Methods}

We perform a theoretical analysis using the density functional theory 
with and without spin-orbit coupling. We use the plane-wave pseudopotential 
method implemented in the Vienna ab-initio simulation package 
(VASP)\cite{Kresse1996} within the generalized gradient approximations (GGA), 
and it gives structural properties closer to the experimental values. The electronic valence considered are: Ru: 5$s^{1}$4$d^{7}$ and O: 2$s^{2}$2$p^{4}$, and Ca: 3$s$3$p$4$s$.
We use a plane-wave energy cutoff of 650 eV and set a Regular Monkhorst-Pack grid of 7$\times$7$\times$5 to perform the atomic relaxation and  11$\times$11$\times$7 to perform the self-consistent calculation. We use a fine k-grid of 14$\times$14$\times$7 within the tetrahedron method for the density of states.
We perform the structural optimization of the unit cell until a force convergence threshold of at least 10$^{-3}$ eV/\AA\ per atom.

To account for electronic correlations in the Ru $d$-orbitals, we use the Dudarev approach \cite{dudarev1998electron} with an effective Hubbard parameter $U = 1.0$ eV, previously shown to yield accurate structural and electronic properties for CRO \cite{leon20243}. \bluemark{We use a Cartesian coordinate system with crystallographic axes aligned as $\mathbf{a} \parallel \hat{x}$, $\mathbf{b} \parallel \hat{y}$, and $\mathbf{c} \parallel \hat{z}$.}

\section*{Acknowledgments}
A.L. thanks the Technische Universität Dresden Professor Fellowship Program.
C. A. acknowledges the useful discussion with G. Cuono and A. Fakhredine about altermagnetism in Ca$_2$RuO$_4$.
C. A. is supported by the Foundation for Polish Science project “MagTop” no. FENG.02.01-IP.05-0028/23 co-financed by the European Union from the funds of Priority 2 of the European Funds for a Smart Economy Program 2021-2027 (FENG). J.W.G. acknowledges financial support from Chilean ANID-FONDECY grants N. 1220700 and 1221301. T.B. would like to thank the German Science Foundation (DFG) for supporting this work via the German-Israeli Project Cooperation with the Grant No. HE 3543/42-1.

\section*{Author Contributions}

A.L. and J.G. performed the calculations and contributed to the organization of the article. A.L. and C.A. held foundational discussions about the study. A.L., J.G., and C.A. carried out the symmetry analysis. A.L. and T.B. performed the electronic structure analysis. A.L., J.G., and C.A. wrote the initial draft. All authors contributed to the discussion of the results and collaborated in editing and shaping the final version of the manuscript.

\section{Competing interests}
The authors declare no competing interests.

\bibliographystyle{unsrtnat}
\bibliography{bib}

\begin{thebibliography}{51}
\providecommand{\natexlab}[1]{#1}
\providecommand{\url}[1]{\texttt{#1}}
\expandafter\ifx\csname urlstyle\endcsname\relax
  \providecommand{\doi}[1]{doi: #1}\else
  \providecommand{\doi}{doi: \begingroup \urlstyle{rm}\Url}\fi

\bibitem[{\v{S}}mejkal et~al.(2022){\v{S}}mejkal, Sinova, and Jungwirth]{vsmejkal2022beyond}
Libor {\v{S}}mejkal, Jairo Sinova, and Tomas Jungwirth.
\newblock Beyond conventional ferromagnetism and antiferromagnetism: A phase with nonrelativistic spin and crystal rotation symmetry.
\newblock \emph{Physical Review X}, 12\penalty0 (3):\penalty0 031042, 2022.

\bibitem[Bai et~al.(2024)Bai, Feng, Liu, {\v{S}}mejkal, Mokrousov, and Yao]{bai2024altermagnetism}
Ling Bai, Wanxiang Feng, Siyuan Liu, Libor {\v{S}}mejkal, Yuriy Mokrousov, and Yugui Yao.
\newblock Altermagnetism: Exploring new frontiers in magnetism and spintronics.
\newblock \emph{arXiv preprint arXiv:2406.02123}, 2024.

\bibitem[Sato et~al.(2024)Sato, Haddad, Fulga, Assaad, and van~den Brink]{sato2024altermagnetic}
Toshihiro Sato, Sonia Haddad, Ion~Cosma Fulga, Fakher~F Assaad, and Jeroen van~den Brink.
\newblock Altermagnetic anomalous hall effect emerging from electronic correlations.
\newblock \emph{Physical Review Letters}, 133\penalty0 (8):\penalty0 086503, 2024.

\bibitem[Chakraborty et~al.(2024{\natexlab{a}})Chakraborty, Gonz\'alez~Hern\'andez, \ifmmode~\check{S}\else \v{S}\fi{}mejkal, and Sinova]{PhysRevB.109.144421}
Atasi Chakraborty, Rafael Gonz\'alez~Hern\'andez, Libor \ifmmode~\check{S}\else \v{S}\fi{}mejkal, and Jairo Sinova.
\newblock Strain-induced phase transition from antiferromagnet to altermagnet.
\newblock \emph{Physical Review B}, 109:\penalty0 144421, Apr 2024{\natexlab{a}}.

\bibitem[Takahashi et~al.(2025)Takahashi, Steward, Ogata, Fernandes, and Schmalian]{takahashi2025elasto}
Keigo Takahashi, Charles~RW Steward, Masao Ogata, Rafael~M Fernandes, and J{\"o}rg Schmalian.
\newblock Elasto-hall conductivity and the anomalous hall effect in altermagnets.
\newblock \emph{arXiv preprint arXiv:2502.03517}, 2025.

\bibitem[Dzyaloshinsky(1958)]{DZYALOSHINSKY1958241}
I.~Dzyaloshinsky.
\newblock A thermodynamic theory of “weak” ferromagnetism of antiferromagnetics.
\newblock \emph{Journal of Physics and Chemistry of Solids}, 4\penalty0 (4):\penalty0 241--255, 1958.

\bibitem[Autieri et~al.(2025{\natexlab{a}})Autieri, Sattigeri, Cuono, and Fakhredine]{autieri2024staggereddzyaloshinskiimoriyainducingweak}
Carmine Autieri, Raghottam~M. Sattigeri, Giuseppe Cuono, and Amar Fakhredine.
\newblock Staggered dzyaloshinskii-moriya interaction inducing weak ferromagnetism in centrosymmetric altermagnets and weak ferrimagnetism in noncentrosymmetric altermagnets.
\newblock \emph{Physical Review B}, 111:\penalty0 054442, 2025{\natexlab{a}}.

\bibitem[Roig et~al.(2024)Roig, Yu, Ekman, Kreisel, Andersen, and Agterberg]{roig2024quasisymmetryconstrainedspinferromagnetism}
Merc{\`e} Roig, Yue Yu, Rune~C Ekman, Andreas Kreisel, Brian~M Andersen, and Daniel~F Agterberg.
\newblock Quasi-symmetry constrained spin ferromagnetism in altermagnets.
\newblock \emph{arXiv preprint arXiv:2412.09338}, 2024.

\bibitem[Cheong and Huang(2024{\natexlab{a}})]{cheong2025altermagnetismclassification}
Sang-Wook Cheong and Fei-Ting Huang.
\newblock Altermagnetism classification.
\newblock \emph{arXiv preprint arXiv:2409.20456}, 2024{\natexlab{a}}.

\bibitem[Cheong and Huang(2024{\natexlab{b}})]{Cheong2024}
Sang-Wook Cheong and Fei-Ting Huang.
\newblock Altermagnetism with non-collinear spins.
\newblock \emph{npj Quantum Materials}, 9\penalty0 (1):\penalty0 13, 2024{\natexlab{b}}.

\bibitem[Kluczyk et~al.(2024)Kluczyk, Gas, Grzybowski, Skupinski, Borysiewicz, Fas, Suffczynski, Domagala, Grasza, Mycielski, Baj, Ahn, V\'yborn\'y, Sawicki, and Gryglas-Borysiewicz]{PhysRevB.110.155201}
K.~P. Kluczyk, K.~Gas, M.~J. Grzybowski, P.~Skupinski, M.~A. Borysiewicz, T.~Fas, J.~Suffczynski, J.~Z. Domagala, K.~Grasza, A.~Mycielski, M.~Baj, K.~H. Ahn, K.~V\'yborn\'y, M.~Sawicki, and M.~Gryglas-Borysiewicz.
\newblock Coexistence of anomalous hall effect and weak magnetization in a nominally collinear antiferromagnet {MnTe}.
\newblock \emph{Physical Review B}, 110:\penalty0 155201, Oct 2024.

\bibitem[Jo et~al.(2024)Jo, Go, Mokrousov, Oppeneer, Cheong, and Lee]{jo2025weakferromagnetismaltermagnetsalternating}
Daegeun Jo, Dongwook Go, Yuriy Mokrousov, Peter~M Oppeneer, Sang-Wook Cheong, and Hyun-Woo Lee.
\newblock Weak ferromagnetism in altermagnets from alternating $ g $-tensor anisotropy.
\newblock \emph{arXiv preprint arXiv:2410.17386}, 2024.

\bibitem[Bernardini et~al.(2025{\natexlab{a}})Bernardini, Fiebig, and Cano]{bernardini2025ruddlesden}
Fabio Bernardini, Manfred Fiebig, and Andr{\'e}s Cano.
\newblock Ruddlesden--popper and perovskite phases as a material platform for altermagnetism.
\newblock \emph{Journal of Applied Physics}, 137\penalty0 (10), 2025{\natexlab{a}}.

\bibitem[Autieri et~al.(2025{\natexlab{b}})Autieri, Cuono, Chakraborty, Gentile, and Black-Schaffer]{autieri2025conditionsorbitalselectivealtermagnetismsr2ruo4}
Carmine Autieri, Giuseppe Cuono, Debmalya Chakraborty, Paola Gentile, and Annica~M Black-Schaffer.
\newblock Conditions for orbital-selective altermagnetism in {Sr$_2$RuO$_4$}: tight-binding model, similarities with cuprates, and implications for superconductivity.
\newblock \emph{arXiv preprint arXiv:2501.14378}, 2025{\natexlab{b}}.

\bibitem[Naka et~al.(2025)Naka, Motome, and Seo]{naka2025altermagnetic}
Makoto Naka, Yukitoshi Motome, and Hitoshi Seo.
\newblock Altermagnetic perovskites.
\newblock \emph{npj Spintronics}, 3\penalty0 (1):\penalty0 1, 2025.

\bibitem[Bernardini et~al.(2025{\natexlab{b}})Bernardini, Fiebig, and Cano]{bernardini2024ruddlesden}
Fabio Bernardini, Manfred Fiebig, and Andrés Cano.
\newblock Ruddlesden–popper and perovskite phases as a material platform for altermagnetism.
\newblock \emph{Journal of Applied Physics}, 137\penalty0 (10):\penalty0 103903, 2025{\natexlab{b}}.

\bibitem[Li et~al.(2024)Li, Leeb, Wohlfeld, Valent{\'\i}, and Knolle]{li2024dwavemagnetismcupratesoxygen}
Ying Li, Valentin Leeb, Krzysztof Wohlfeld, Roser Valent{\'\i}, and Johannes Knolle.
\newblock d-wave magnetism in cuprates from oxygen moments.
\newblock \emph{arXiv preprint arXiv:2412.11922}, 2024.

\bibitem[Ricco et~al.(2018)Ricco, Kim, Tamai, McKeown~Walker, Bruno, Cucchi, Cappelli, Besnard, Kim, Dudin, et~al.]{ricco2018situ}
Sara Ricco, Minjae Kim, Anna Tamai, S~McKeown~Walker, Flavio~Yair Bruno, Irene Cucchi, Edoardo Cappelli, C{\'e}line Besnard, Timur~K Kim, Pavel Dudin, et~al.
\newblock In situ strain tuning of the metal-insulator-transition of {Ca$_{2}$RuO$_{4}$} in angle-resolved photoemission experiments.
\newblock \emph{Nature Communications}, 9\penalty0 (1):\penalty0 4535, 2018.

\bibitem[Tiwari et~al.(2023)Tiwari, He, Fu, Sun, Perry, Mackenzie, and Julian]{tiwari2023suppression}
P~Tiwari, L~He, M~Fu, D~Sun, RS~Perry, AP~Mackenzie, and SR~Julian.
\newblock {Suppression of field-induced spin density wave order in Sr$_{3}$Ru$_{2}$O$_{7}$ with pressure}.
\newblock \emph{Physical Review B}, 108\penalty0 (11):\penalty0 115154, 2023.

\bibitem[Maeno et~al.(2024)Maeno, Yonezawa, and Ramires]{maeno2024still}
Yoshiteru Maeno, Shingo Yonezawa, and Aline Ramires.
\newblock {Still Mystery after All These Years—Unconventional Superconductivity of Sr$_{2}$RuO$_{4}$}.
\newblock \emph{Journal of the Physical Society of Japan}, 93\penalty0 (6):\penalty0 062001, 2024.

\bibitem[Yoshida et~al.(2005)Yoshida, Ikeda, Matsuhata, Shirakawa, Lee, and Katano]{yoshida2005crystal}
Yoshiyuki Yoshida, Shin-Ichi Ikeda, Hirofumi Matsuhata, Naoki Shirakawa, CH~Lee, and Susumu Katano.
\newblock {Crystal and magnetic structure of {Ca$_{3}$Ru$_{2}$O$_{7}$}}.
\newblock \emph{Physical Review B}, 72\penalty0 (5):\penalty0 054412, 2005.

\bibitem[Kikugawa et~al.(2007)Kikugawa, Rost, Baumberger, Ingle, Hossain, Meevasana, Shen, Lu, Damascelli, Mackenzie, et~al.]{kikugawa2007ca3ru2o7}
N~Kikugawa, A~Rost, F~Baumberger, NJC Ingle, MA~Hossain, W~Meevasana, KM~Shen, DH~Lu, A~Damascelli, AP~Mackenzie, et~al.
\newblock {Ca$_{3}$Ru$_{2}$O$_{7}$: Electronic instability and extremely strong quasiparticle renormalisation}.
\newblock \emph{Journal of magnetism and magnetic materials}, 310\penalty0 (2):\penalty0 1027--1029, 2007.

\bibitem[von Arx et~al.(2020)von Arx, Forte, Horio, Granata, Wang, Das, Sassa, Fittipaldi, Fatuzzo, Ivashko, Tseng, Paris, Vecchione, Schmitt, Cuoco, and Chang]{PhysRevB.102.235104}
K.~von Arx, F.~Forte, M.~Horio, V.~Granata, Q.~Wang, L.~Das, Y.~Sassa, R.~Fittipaldi, C.~G. Fatuzzo, O.~Ivashko, Y.~Tseng, E.~Paris, A.~Vecchione, T.~Schmitt, M.~Cuoco, and J.~Chang.
\newblock Resonant inelastic x-ray scattering study of ${\mathrm{ca}}_{3}{\mathrm{ru}}_{2}{\mathrm{o}}_{7}$.
\newblock \emph{Physical Review B}, 102:\penalty0 235104, 2020.

\bibitem[Ohmichi et~al.(2004)Ohmichi, Yoshida, Ikeda, Shirakawa, and Osada]{ohmichi2004colossal}
E~Ohmichi, Y~Yoshida, SI~Ikeda, N~Shirakawa, and T~Osada.
\newblock {Colossal magnetoresistance accompanying a structural transition in a highly two-dimensional metallic state of Ca$_{3}$Ru$_{2}$O$_{7}$}.
\newblock \emph{Physical Review B}, 70\penalty0 (10):\penalty0 104414, 2004.

\bibitem[Karpus et~al.(2006)Karpus, Snow, Gupta, Barath, Cooper, and Cao]{karpus2006spectroscopic}
JF~Karpus, CS~Snow, R~Gupta, H~Barath, SL~Cooper, and G~Cao.
\newblock {Spectroscopic study of the field-and pressure-induced phases of the bilayered ruthenate Ca$_{3}$Ru$_{2}$O$_{7}$}.
\newblock \emph{Physical Review B}, 73\penalty0 (13):\penalty0 134407, 2006.

\bibitem[Dashwood et~al.(2023)Dashwood, Walker, Kwasigroch, Veiga, Faure, Vale, Porter, Manuel, Khalyavin, Orlandi, et~al.]{dashwood2023strain}
CD~Dashwood, AH~Walker, MP~Kwasigroch, LSI Veiga, Q~Faure, JG~Vale, DG~Porter, P~Manuel, DD~Khalyavin, F~Orlandi, et~al.
\newblock {Strain control of a bandwidth-driven spin reorientation in Ca$_{3}$Ru$_{2}$O$_{7}$}.
\newblock \emph{Nature Communications}, 14\penalty0 (1):\penalty0 6197, 2023.

\bibitem[Markovi{\'c} et~al.(2020)Markovi{\'c}, Watson, Clark, Mazzola, Abarca~Morales, Hooley, Rosner, Polley, Balasubramanian, Mukherjee, et~al.]{markovic2020electronically}
Igor Markovi{\'c}, Matthew~D Watson, Oliver~J Clark, Federico Mazzola, Edgar Abarca~Morales, Chris~A Hooley, Helge Rosner, Craig~M Polley, Thiagarajan Balasubramanian, Saumya Mukherjee, et~al.
\newblock {Electronically driven spin-reorientation transition of the correlated polar metal Ca$_3$Ru$_2$O$_7$}.
\newblock \emph{Proceedings of the National Academy of Sciences}, 117\penalty0 (27):\penalty0 15524--15529, 2020.

\bibitem[{\v{S}}mejkal(2024)]{smejkal2024altermagneticmultiferroicsaltermagnetoelectriceffect}
Libor {\v{S}}mejkal.
\newblock Altermagnetic multiferroics and altermagnetoelectric effect.
\newblock \emph{arXiv preprint arXiv:2411.19928}, 2024.

\bibitem[Grzybowski et~al.(2024)Grzybowski, Autieri, Domagala, Krasucki, Kaleta, Kret, Gas, Sawicki, Bożek, Suffczyński, and Pacuski]{D3NR04798A}
Michał~J. Grzybowski, Carmine Autieri, Jaroslaw Domagala, Cezary Krasucki, Anna Kaleta, Sławomir Kret, Katarzyna Gas, Maciej Sawicki, Rafał Bożek, Jan Suffczyński, and Wojciech Pacuski.
\newblock {Wurtzite vs. rock-salt MnSe epitaxy: electronic and altermagnetic properties}.
\newblock \emph{Nanoscale}, 16:\penalty0 6259--6267, 2024.

\bibitem[Schiff et~al.(2024)Schiff, McClarty, Rau, and Romhanyi]{schiff2024collinear}
Hana Schiff, Paul McClarty, Jeffrey~G Rau, and Judit Romhanyi.
\newblock Collinear altermagnets and their landau theories.
\newblock \emph{arXiv preprint arXiv:2412.18025}, 2024.

\bibitem[Yuan et~al.(2023)Yuan, Zhang, Acosta, and Zunger]{Yuan2023}
Lin-Ding Yuan, Xiuwen Zhang, Carlos~Mera Acosta, and Alex Zunger.
\newblock Uncovering spin-orbit coupling-independent hidden spin polarization of energy bands in antiferromagnets.
\newblock \emph{Nature Communications}, 14\penalty0 (1):\penalty0 5301, Aug 2023.

\bibitem[Gallego et~al.(2012)Gallego, Tasci, Flor, Perez-Mato, and Aroyo]{gallego2012magnetic}
Samuel~V Gallego, Emre~S Tasci, G~Flor, J~Manuel Perez-Mato, and Mois~I Aroyo.
\newblock Magnetic symmetry in the bilbao crystallographic server: a computer program to provide systematic absences of magnetic neutron diffraction.
\newblock \emph{Journal of Applied Crystallography}, 45\penalty0 (6):\penalty0 1236--1247, 2012.

\bibitem[Ke et~al.(2011)Ke, Peng, Singh, Hong, Tian, Dela~Cruz, and Mao]{ke2011emergent}
Xianglin Ke, J~Peng, DJ~Singh, Tao Hong, Wei Tian, CR~Dela~Cruz, and ZQ~Mao.
\newblock {Emergent electronic and magnetic state in Ca$_{3}$Ru$_{2}$O$_{7}$ induced by Ti doping}.
\newblock \emph{Physical Review B}, 84\penalty0 (20):\penalty0 201102, 2011.

\bibitem[Rivero et~al.(2017)Rivero, Jin, Chen, Meunier, Plummer, and Shelton]{rivero2017predicting}
Pablo Rivero, Rongying Jin, Chen Chen, Vincent Meunier, EW~Plummer, and William Shelton.
\newblock {Predicting hidden bulk phases from surface phases in bilayered Sr$_{3}$Ru$_{2}$O$_{7}$}.
\newblock \emph{Scientific Reports}, 7\penalty0 (1):\penalty0 10265, 2017.

\bibitem[Le{\'o}n et~al.(2024)Le{\'o}n, Gonz{\'a}lez, and Rosner]{leon20243}
AM~Le{\'o}n, JW~Gonz{\'a}lez, and Helge Rosner.
\newblock {Ca$_{3}$Ru$_{2}$O$_{7}$: Interplay among degrees of freedom and the role of the exchange correlation}.
\newblock \emph{Physical Review Materials}, 8\penalty0 (2):\penalty0 024411, 2024.

\bibitem[Gonz{\'a}lez et~al.(2025)Gonz{\'a}lez, Le{\'o}n, Gonz{\'a}lez-Fuentes, and Gallardo]{gonzalez2025altermagnetism}
JW~Gonz{\'a}lez, AM~Le{\'o}n, C~Gonz{\'a}lez-Fuentes, and RA~Gallardo.
\newblock {Altermagnetism in two-dimensional Ca$_{2}$RuO$_{4}$ perovskite}.
\newblock \emph{Nanoscale}, 17\penalty0 (8):\penalty0 4796--4807, 2025.

\bibitem[Puggioni et~al.(2020)Puggioni, Horio, Chang, and Rondinelli]{puggioni2020cooperative}
Danilo Puggioni, M~Horio, J~Chang, and James~M Rondinelli.
\newblock {Cooperative interactions govern the fermiology of the polar metal Ca$_{3}$Ru$_{2}$O$_{7}$}.
\newblock \emph{Physical Review Research}, 2\penalty0 (2):\penalty0 023141, 2020.

\bibitem[Liu(2011)]{liu2011mott}
Guo-Qiang Liu.
\newblock {Mott transition and magnetic anisotropy in Ca$_3$Ru$_2$O$_7$}.
\newblock \emph{Physical Review B}, 84\penalty0 (23):\penalty0 235137, 2011.

\bibitem[Cuono et~al.(2023)Cuono, Sattigeri, Skolimowski, and Autieri]{cuono2023orbital}
Giuseppe Cuono, Raghottam~M Sattigeri, Jan Skolimowski, and Carmine Autieri.
\newblock Orbital-selective altermagnetism and correlation-enhanced spin-splitting in strongly-correlated transition metal oxides.
\newblock \emph{Journal of Magnetism and Magnetic Materials}, 586:\penalty0 171163, 2023.

\bibitem[Horio et~al.(2021)Horio, Wang, Granata, Kramer, Sassa, J{\"o}hr, Sutter, Bold, Das, Xu, et~al.]{horio2021electronic}
M~Horio, Q~Wang, V~Granata, KP~Kramer, Y~Sassa, S~J{\"o}hr, D~Sutter, A~Bold, L~Das, Y~Xu, et~al.
\newblock {Electronic reconstruction forming a C 2-symmetric Dirac semimetal in Ca$_{3}$Ru$_{2}$O$_{7}$}.
\newblock \emph{npj Quantum Materials}, 6\penalty0 (1):\penalty0 29, 2021.

\bibitem[Cheong and Huang(2024{\natexlab{c}})]{cheong2024altermagnetism}
Sang-Wook Cheong and Fei-Ting Huang.
\newblock Altermagnetism with non-collinear spins.
\newblock \emph{npj Quantum Materials}, 9\penalty0 (1):\penalty0 13, 2024{\natexlab{c}}.

\bibitem[{\v{S}}mejkal()]{vsmejkal2411altermagnetic}
L~{\v{S}}mejkal.
\newblock Altermagnetic multiferroics and altermagnetoelectric effect (2024).
\newblock \emph{arXiv preprint arXiv:2411.19928}.

\bibitem[Bhowal and Spaldin(2024)]{bhowal2024ferroically}
Sayantika Bhowal and Nicola~A Spaldin.
\newblock Ferroically ordered magnetic octupoles in d-wave altermagnets.
\newblock \emph{Physical Review X}, 14\penalty0 (1):\penalty0 011019, 2024.

\bibitem[Wang et~al.(2024)Wang, Bordoloi, Ding, Wang, Wu, Lin, Yang, Liu, Zhou, Meng, et~al.]{wang2024transport}
Zhen Wang, Arjyama Bordoloi, Zhaoqing Ding, Enling Wang, Xiaofeng Wu, Zeguo Lin, Mingyu Yang, Chenxu Liu, Jinglin Zhou, Meng Meng, et~al.
\newblock {Transport and magnetic properties of Hund's metal CaRuO$_{3}$ under strain modulation}.
\newblock \emph{Physical Review B}, 110\penalty0 (4):\penalty0 L041403, 2024.

\bibitem[Burganov et~al.(2016)Burganov, Adamo, Mulder, Uchida, King, Harter, Shai, Gibbs, Mackenzie, Uecker, et~al.]{burganov2016strain}
Bulat Burganov, Carolina Adamo, Andrew Mulder, M~Uchida, PDC King, JW~Harter, DE~Shai, AS~Gibbs, AP~Mackenzie, Reinhard Uecker, et~al.
\newblock Strain control of fermiology and many-body interactions in two-dimensional ruthenates.
\newblock \emph{Physical review letters}, 116\penalty0 (19):\penalty0 197003, 2016.

\bibitem[Dong et~al.(2024)Dong, Zhang, Cao, Chen, Lu, Wang, and Wu]{dong2024polar}
Mingdong Dong, Yichi Zhang, Jing-ming Cao, Haowen Chen, Qiyang Lu, Hong-fei Wang, and Jie Wu.
\newblock Polar metallicity controlled by epitaxial strain engineering.
\newblock \emph{Advanced Science}, 11\penalty0 (40):\penalty0 2408329, 2024.

\bibitem[Tsubaki et~al.(2023)Tsubaki, Arita, Katase, Kamiya, Tsurumaki-Fukuchi, and Takahashi]{tsubaki2023significant}
Keiji Tsubaki, Masashi Arita, Takayoshi Katase, Toshio Kamiya, Atsushi Tsurumaki-Fukuchi, and Yasuo Takahashi.
\newblock Significant effects of epitaxial strain on the nonlinear transport properties in ca2ruo4 thin films with the current-driven transition.
\newblock \emph{Japanese Journal of Applied Physics}, 63\penalty0 (1):\penalty0 01SP03, 2023.

\bibitem[Chakraborty et~al.(2024{\natexlab{b}})Chakraborty, Gonz{\'a}lez~Hern{\'a}ndez, {\v{S}}mejkal, and Sinova]{chakraborty2024strain}
Atasi Chakraborty, Rafael Gonz{\'a}lez~Hern{\'a}ndez, Libor {\v{S}}mejkal, and Jairo Sinova.
\newblock Strain-induced phase transition from antiferromagnet to altermagnet.
\newblock \emph{Physical Review B}, 109\penalty0 (14):\penalty0 144421, 2024{\natexlab{b}}.

\bibitem[Xun et~al.(2025)Xun, Liu, Zhang, Wu, and Li]{xun2025stacking}
Wei Xun, Xin Liu, Youdong Zhang, Yin-Zhong Wu, and Ping Li.
\newblock Stacking-, strain-engineering induced altermagnetism, multipiezo effect, and topological state in two-dimensional materials.
\newblock \emph{Applied Physics Letters}, 126\penalty0 (16), 2025.

\bibitem[Kresse and Furthm{\"u}ller(1996)]{Kresse1996}
Georg Kresse and J{\"u}rgen Furthm{\"u}ller.
\newblock Efficiency of ab-initio total energy calculations for metals and semiconductors using a plane-wave basis set.
\newblock \emph{Computational Materials Science}, 6\penalty0 (1):\penalty0 15--50, 1996.

\bibitem[Dudarev et~al.(1998)Dudarev, Botton, Savrasov, Humphreys, and Sutton]{dudarev1998electron}
Sergei~L Dudarev, Gianluigi~A Botton, Sergey~Y Savrasov, CJ~Humphreys, and Adrian~P Sutton.
\newblock {Electron-energy-loss spectra and the structural stability of nickel oxide: An LSDA+U study}.
\newblock \emph{Physical Review B}, 57\penalty0 (3):\penalty0 1505, 1998.

\end{thebibliography}


\clearpage

\begin{thebibliography}{99}

\bibitem{yoshida2005}
Y. Yoshida, S.-I. Ikeda, H. Matsuhata, N. Shirakawa, C.-H. Lee, and S. Katano, 
Crystal and magnetic structure of Ca$_{3}$Ru$_{2}$O$_{7}$, 
\textit{Phys. Rev. B} \textbf{72}, 054412 (2005).

\bibitem{puggioni2020}
D. Puggioni, M. Horio, J. Chang, and J. M. Rondinelli, 
Cooperative interactions govern the fermiology of the polar metal Ca$_{3}$Ru$_{2}$O$_{7}$, 
\textit{Phys. Rev. Res.} \textbf{2}, 023141 (2020).

\bibitem{leon2024}
A. M. León, J. W. Gonzalez, and H. Rosner, 
Ca$_{3}$Ru$_{2}$O$_{7}$: Interplay among degrees of freedom and the role of the exchange correlation, 
\textit{Phys. Rev. Mater.} \textbf{8}, 024411 (2024).

\bibitem{fukaya2025}
Y. Fukaya, B. Lu, K. Yada, Y. Tanaka, and J. Cayao, 
Superconducting phenomena in systems with unconventional magnets, 
arXiv:2502.15400 (2025).

\bibitem{hayami2020}
S. Hayami, Y. Yanagi, and H. Kusunose, 
Bottom-up design of spin-split and reshaped electronic band structures in antiferromagnets without spin–orbit coupling: Procedure on the basis of augmented multipoles, 
\textit{Phys. Rev. B} \textbf{102}, 144441 (2020).


\end{thebibliography}
\newpage
\pagebreak

\onecolumn
\begin{center}

\section*{Supplementary Information}

\textbf{\large{Hybrid d/p-wave altermagnetism in Ca$_3$Ru$_2$O$_7$ and strain-controlled spin splitting}}

\end{center}

\setcounter{figure}{0} 
\setcounter{section}{0} 
\setcounter{equation}{0}
\setcounter{table}{0}
\setcounter{page}{1}
\renewcommand{\thepage}{S\arabic{page}} 
\renewcommand{\thesection}{S\Roman{section}}   
\renewcommand{\thetable}{S\arabic{table}}  
\renewcommand{\thefigure}{S\arabic{figure}} 
\renewcommand{\theequation}{S\arabic{equation}} 

\section{Ca$_{3}$Ru$_{2}$O$_{7}$ structural parameters}

Table \ref{table:S1} shows the energy difference regarding the ground state system (Configuration A) considering spin orbit coupling (SOC) and without SOC (NSOC), respectively (considering the spins along $b$ direction). $a$,$b$ and $c$ are the lattice parameter (obtained from NSOC calculations). 

\begin{table}[h!]
\centering
    \begin{tabular}{c|c|c|c|c}
        \hline
        \textbf{Configuration} & \boldmath{$\Delta$E$_{NSOC}$ (meV/Ru)} &\boldmath{$\Delta$E$_{SOC}$ (meV/Ru)} &\textbf{a/b} (\AA)  &\textbf{c} (\AA)  \\
        \hline
        A & 0   & 0 &  5.415/5.601 & 19.54  \\
        \hline
        B & 34.90 & 36.86 & 5.415/5.726  & 19.23  \\
        \hline
        C & 36.46 & 60.55 & 5.398/5.652 & 19.449 \\
        \hline
        D & 39.22 & 37.32  & 5.419/5.722 & 19.261  \\
        \hline
        Exp \cite{yoshida2005} & -- & -- & 5.367/5.535 & 19.521 \\
        \hline
    \end{tabular}
    \caption{The first column indicates the label corresponding to each magnetic configuration. \textbf{\( \Delta E_{\text{SOC}} \)} and \textbf{\( \Delta E_{\text{NSOC}} \)} represent the energy difference relative to configuration A \((E_{\text{conf.}} - E_{A})\) in meV/Ru, calculated with and without spin-orbit coupling, respectively. \textbf{a/b} and \textbf{c} correspond to the lattice parameters at the equilibrium volume for each configuration.}
    \label{table:S1}
\end{table}

\textbf{Note:} The calculations including spin-orbit coupling  were performed using the relaxed structures obtained from collinear calculations and do not include further structural relaxation.\\

\textbf{Effect Hubbard-U energy differences:}
Table~\ref{table:S2} shows the energy difference relative to the ground state (Configuration A) for three different values of the Hubbard $U$ parameter. We observe that the choice of $U$ does not significantly affect the relative stability of the magnetic phases with respect to configuration A. For $U > 1.2$\,eV, configuration A begins to exhibit structural changes, with distortions in the octahedral environment indicating the onset of structural instability. This behavior has been previously investigated \cite{puggioni2020}, and it has been established that at $U = 1.4$\,eV the system stabilizes in a different structural phase \cite{leon2024,puggioni2020}.

\begin{table}[h!]
\centering
    \begin{tabular}{c|c|c|c}
        \hline
        \textbf{Configuration} & \boldmath{$\Delta$E (U = 0.5)} &\boldmath{$\Delta$E (U = 1.0)} & \boldmath{$\Delta$E (U = 1.1)} \\
        \hline
        A & 0   & 0 &  0  \\
        \hline
        B & 48.41 & 34.90 &  30.76  \\
        \hline
        C & 52.35 & 36.46 &  32.79  \\
        \hline
        D & 74.0 & 39.23 & 37.29   \\
        \hline
        \hline
    \end{tabular}
    \caption{The first column indicates the label corresponding to each magnetic configuration. \textbf{\( \Delta E \)} represents the energy difference relative to configuration A \((E_{\text{conf.}} - E_{A})\), expressed in meV/Ru, and calculated without including spin--orbit coupling. Results are shown for Hubbard-\( U \) values of 0.5, 1.0, and 1.1\,eV.}
    \label{table:S2}
\end{table}

\newpage
\section{Density of states with and without Hubbard-U interaction.}
\label{DOS-U}
\begin{figure}[ht]
\centering
\includegraphics[width=0.3\textwidth]{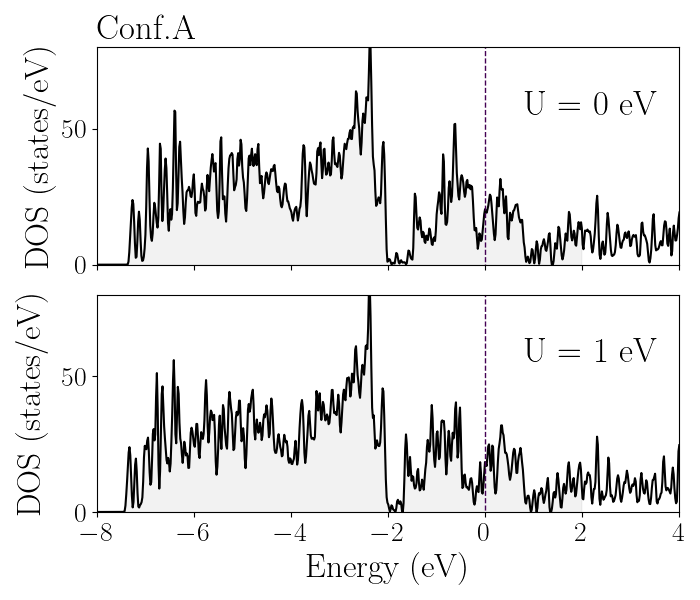}\includegraphics[width=0.3\textwidth]{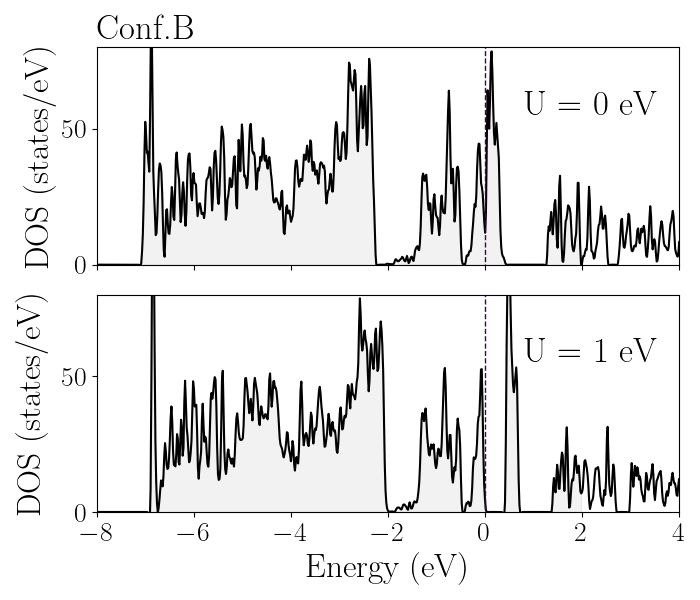}
\caption{Upper and lower panels show the total density of states (DOS) for magnetic configurations A and B, with \( U = 0 \) and 1~eV, respectively.
}
\label{fig:Sdos}
\end{figure}

Figure \ref{fig:Sdos} shows the total density of states (DOS) for configurations A and B, with and without the inclusion of the Hubbard-U interaction. At U=0 eV, both systems exhibit a metallic state. When the Hubbard-
U interaction is introduced, configuration B favors a narrow-gap insulating state, whereas configuration A remains metallic.\\

This different response to the Hubbard-U
 interaction arises from the localization versus delocalization effects associated with the symmetry of the wave functions, which in turn are determined by the magnetic configurations. Specifically, configuration B corresponds to a singlet state, where the spin part of the wave function is antisymmetric under exchange, leading to a symmetric (bonding-like) spatial part that favors electron delocalization. In contrast, configuration A resembles a triplet state, with a symmetric spin part and thus an antisymmetric (antibonding-like) spatial part, resulting in more localized electron clouds.\\

 In essence, the Hubbard-\( U \) term promotes electron localization. Since configuration A already exhibits a relatively localized electronic character, the effect of a small \( U \) value is minimal. However, by increasing \( U \geq 1.4 \, \text{eV} \), a gap opens, leading to an insulating state with a narrow gap, similar to that observed in configuration B. More details about the interplay among the different degrees of freedom in configuration A can be found in Ref.~\cite{leon2024}. In contrast, configuration B initially favors a more delocalized state; therefore, the introduction of \( U \) induces localization, which perturbs the spatial part of the wave function and, consequently, alters the magnetic order.

\section{Ca$_{3}$Ru$_{2}$O$_{7}$ altermagnetic configurations}
\begin{figure*}[ht]
\centering
\begin{tikzpicture}
\node at (2, 0.5) {spin up};
\draw[red, ultra thick] (2,0) -- (3,0);
\node at (5, 0.5) {spin down};
\draw[blue, ultra thick] (4.5,0) -- (5.5,0);
    \end{tikzpicture}
    
\vspace{0.5cm}    
{\includegraphics[width=1.0\textwidth]{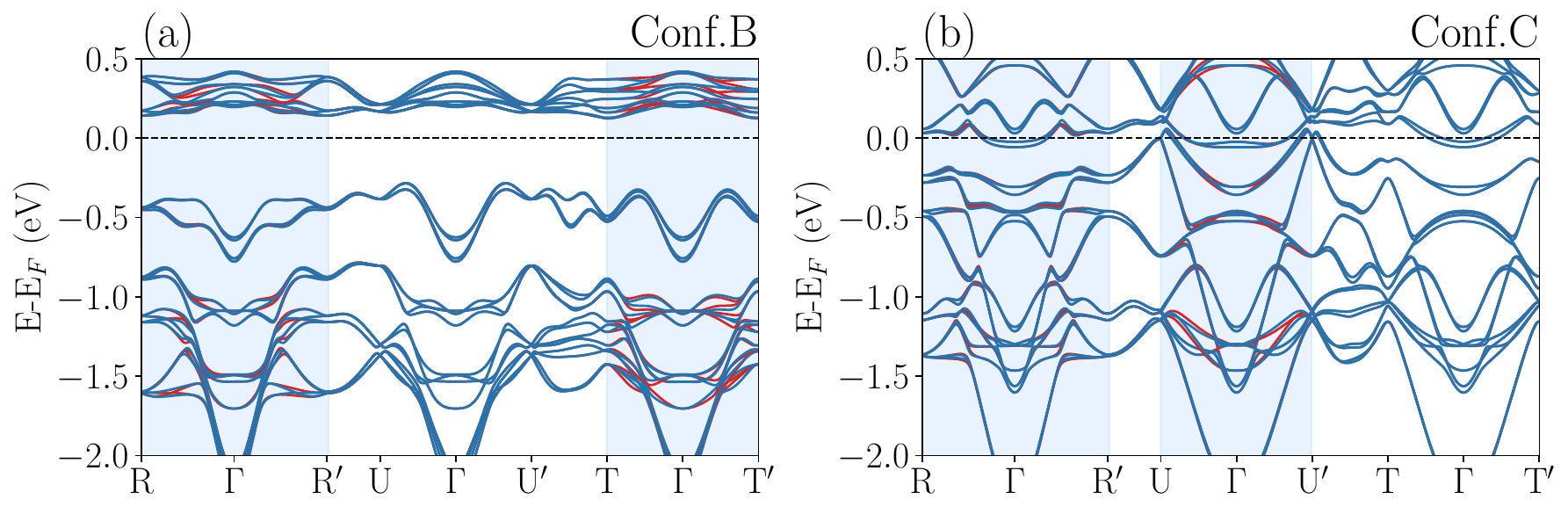}}
\caption{Upper panel: Band structures for configurations B and C.} 
\label{fig:S1}
\end{figure*}

Fig.~\ref{fig:S1} (upper panel) shows the band structure for the AM configurations of Ca$_{3}$Ru$_{2}$O$_{7}$: configuration B (a) and configuration C (b). \\

\begin{figure*}
\centering
\subfigure[]{\includegraphics[width=0.30\textwidth]{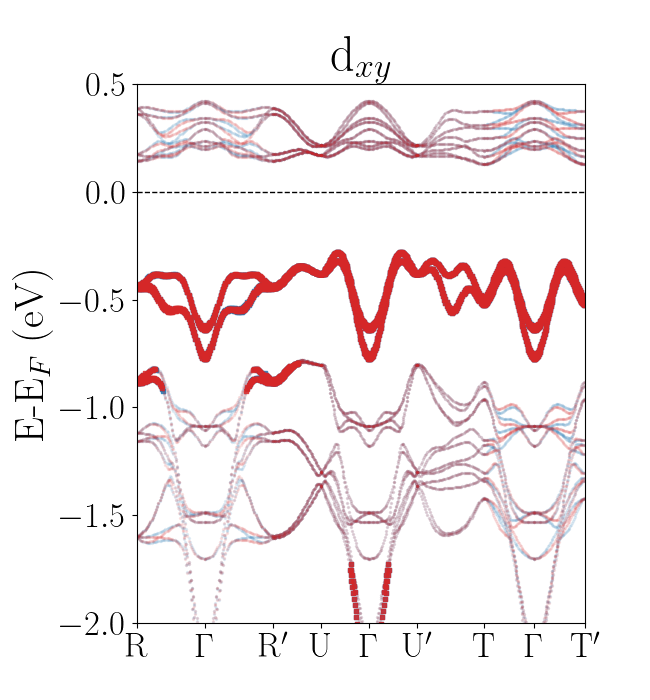}}
\subfigure[]{\includegraphics[width=0.30\textwidth]{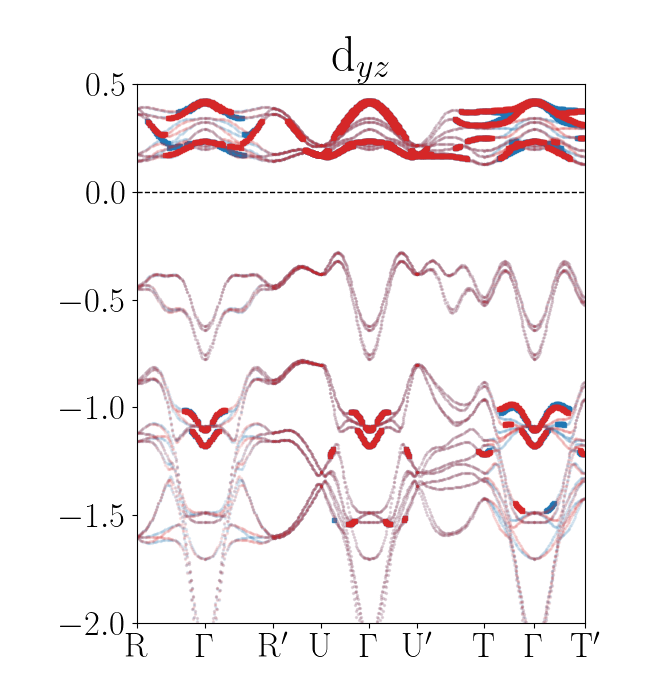}}
\subfigure[]{\includegraphics[width=0.30\textwidth]{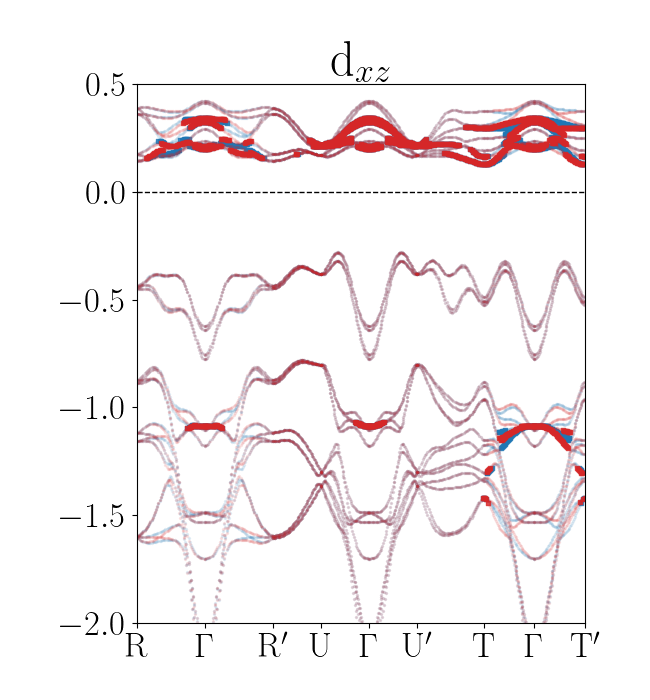}}
\vspace{0.1em} 
\subfigure[]{\includegraphics[width=0.30\textwidth]{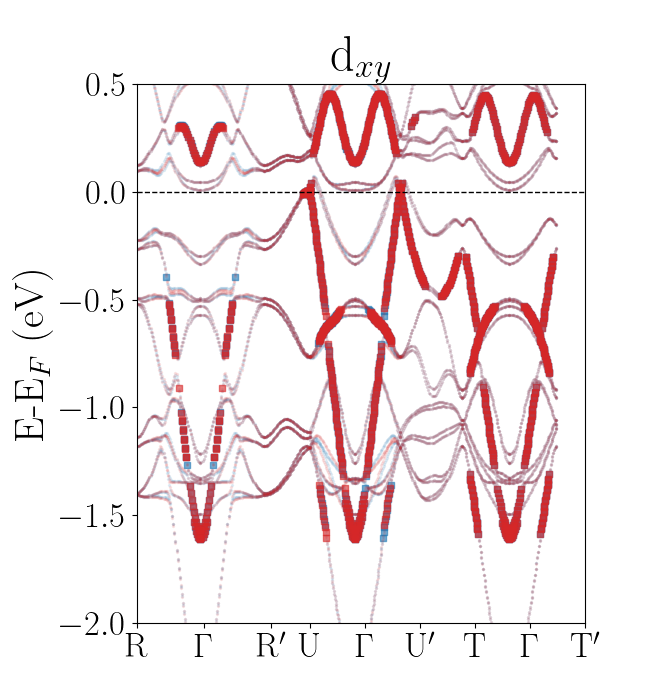}}
\subfigure[]{\includegraphics[width=0.30\textwidth]{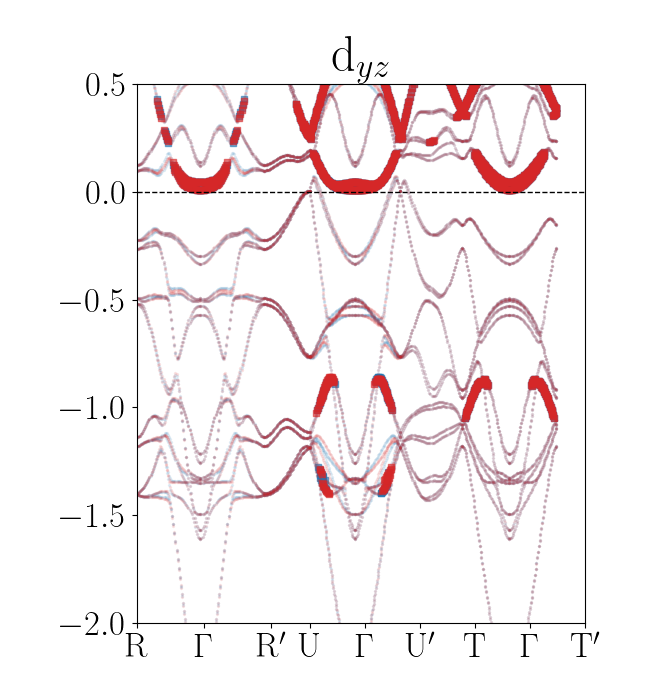}}
\subfigure[]{\includegraphics[width=0.30\textwidth]{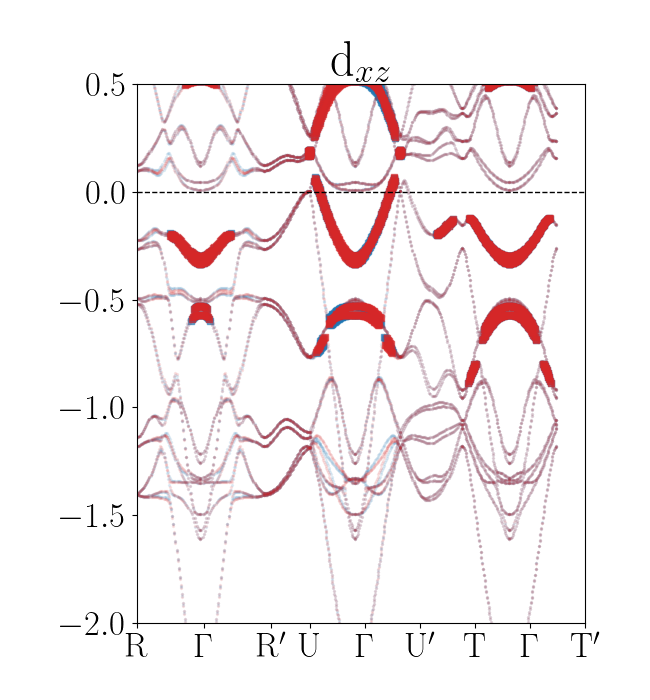}}
\vspace{0.5em}
\begin{tikzpicture}  
    \node at (2, 2) {spin up  $\geq$ 40\% of occupation};
    \node[draw, fill=blue, minimum size=3.5mm] at (2.5, 1.5) {}; 
    \node at (8, 2) {spin down $\geq$ 40\% of occupation};
    \node[draw, fill=red, minimum size=3.5mm] at (8, 1.5) {};    
    \node at (2, 0.5) {spin up  $\leq$ 40\% of occupation};
    \node[draw, fill=blue, minimum size=0.2mm, opacity=0.3] at (2.5, 0) {};  
    \node at (8, 0.5) {spin down $\leq$ 40\% of occupation};
    \node[draw, fill=red, minimum size=0.2mm, opacity=0.3] at (8, 0) {}; 
\end{tikzpicture}
\caption{Band structure projected onto the \( t_{2g} \) orbitals, resolved by occupation. Panels (a)-(c) correspond to configuration B, and panels (d)-(f) to configuration C.}
\label{fig:S2}
\end{figure*}

Fig.~\ref{fig:S2} shows the band structure projected onto the $d$-orbitals by occupation state for the B and C configurations. While the configuration B hosts a P-2 d$_{yz}$-wave altermagnetism, the configuration C hosts a d$_{xz}$-wave altermagnetism. Configuration C exhibits altermagnetism in both d$_{xy}$ and 
d$_{xz}$-d$_{yz}$ bands,  therefore, the altermagnetism of the configuration C is not orbital selective. In the case of the B configuration, the main contributions to the AM bands are between -1.5 and -1.0 eV.\\

\newpage
\section{Relativistic bands}
\label{sec:soc}

\begin{figure*}[h]
  \centering
  \subfigure[]{\includegraphics[width=0.3\textwidth]{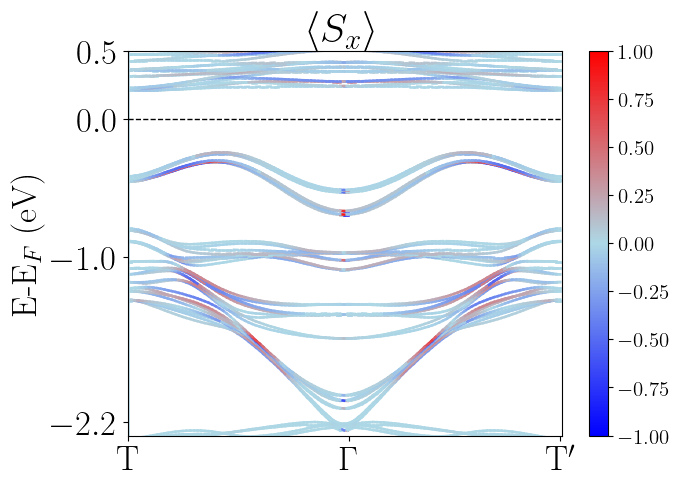}}
  \subfigure[]{\includegraphics[width=0.3\textwidth]{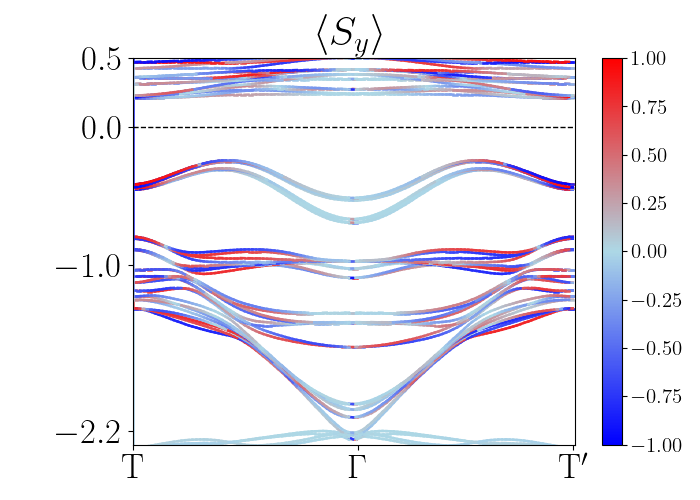}}
  \subfigure[]{\includegraphics[width=0.3\textwidth]{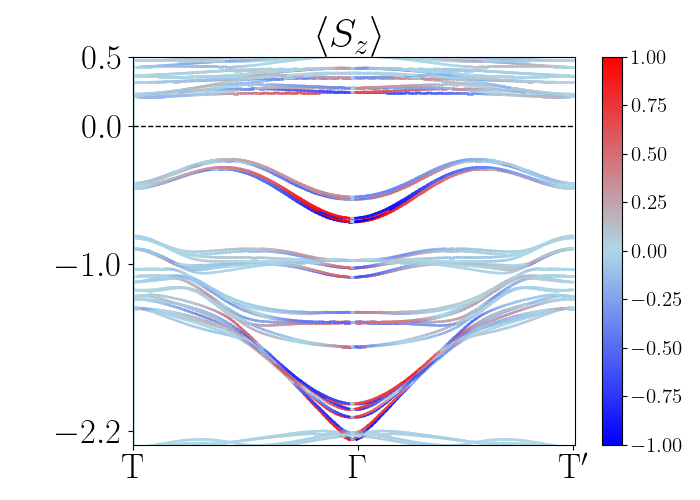}} \\
  \subfigure[]{\includegraphics[width=0.3\textwidth]{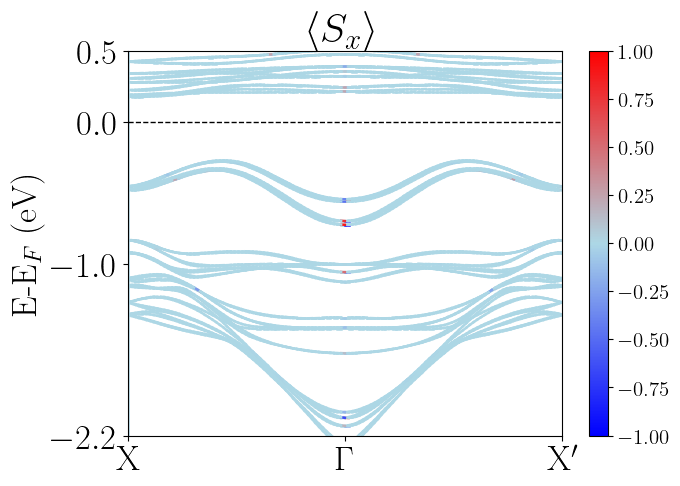}}
  \subfigure[]{\includegraphics[width=0.3\textwidth]{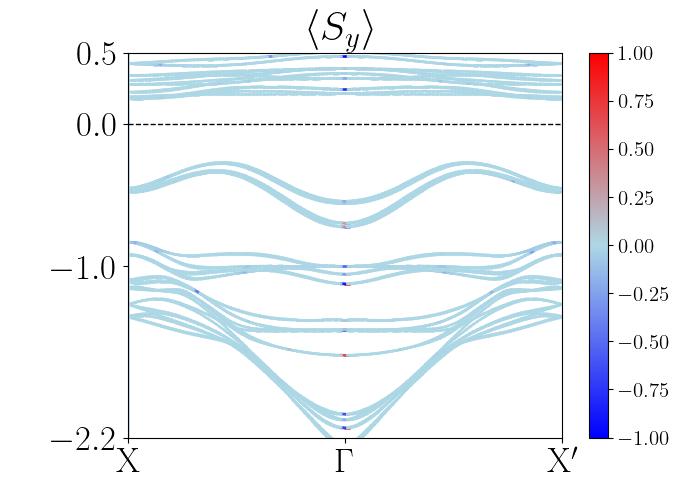}}
  \subfigure[]{\includegraphics[width=0.3\textwidth]{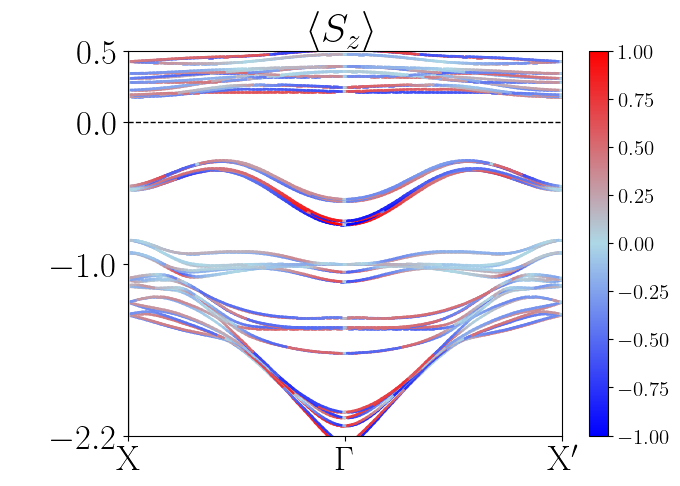}}
  \caption{Projected band structure for configuration B with spin-orbit coupling and the N\'eel vector aligned along the \( b \)-axis. The magnetic moments are projected along the (a)-(d) \( x \)-axis, (b)-(e) \( y \)-axis, and (c)-(f) \( z \)-axis. The color scale represents the expectation value of the corresponding spin component, with blue indicating negative values and red indicating positive values.}
  \label{fig:S4}
\end{figure*}

Figure~\ref{fig:S4} shows the band structure for configuration B. \\

\textbf{Upper panel:} Band structure along the \( \mathrm{T} \rightarrow \Gamma \rightarrow \mathrm{T}' \) path, where altermagnetic  splitting is most pronounced (see Fig.~4 in the main text). The \( S_y \) projection retains strong AM features at lower energies, consistent with the collinear case. In contrast, the \( S_z \) projection reveals AM features between $-0.75$\,eV and $-0.5$\,eV.\\

\textbf{Lower panel:} Band structure along the \( \mathrm{X} \rightarrow \Gamma \rightarrow \mathrm{X}' \) path. The \( S_y \) and \( S_x \) components retain the antiferromagnetic character, while the \( S_z \) projection shows clear AM features near the Fermi level.

\section{Band structure under strain}
\label{sec:amq}
\begin{figure}[ht]
\centering
\includegraphics[width=0.8\textwidth]{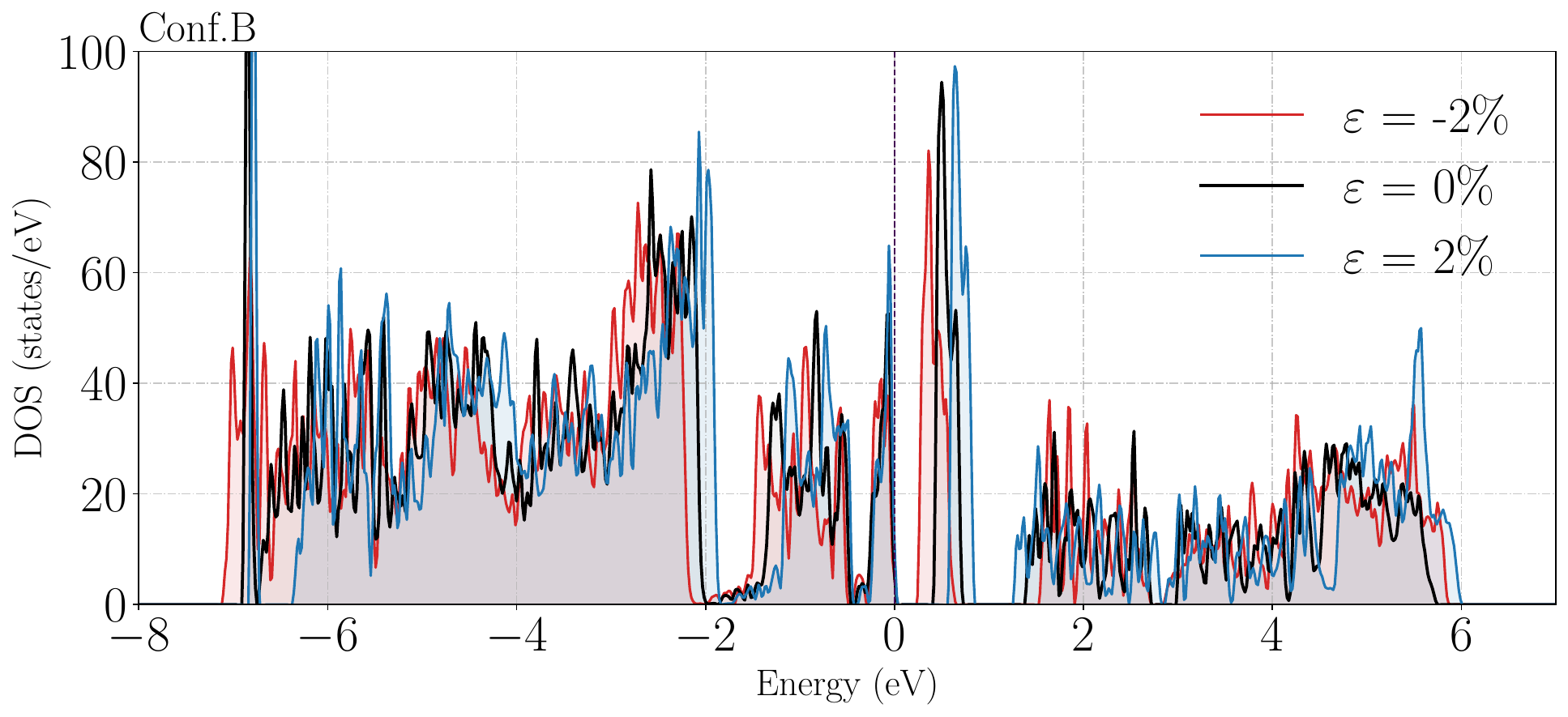}
\caption{Total Density of states (DOS) under three different strain values for B configuration.}
\label{fig:S5}
\end{figure}

Figure~\ref{fig:S4} shows the variation of the DOS at the Fermi level as a function of strain. It can be observed that under compressive strain, the DOS becomes broader, while under tensile strain, it becomes more localized. \\

Figure~\ref{fig:S5} presents the band structure under different strain values ($ab$-plane) for two cases. The upper panels correspond to fully relaxed structures, while the lower panels depict the same strain conditions (same lattice parameters) but with atomic positions fixed at their zero-strain values (non-relaxed calculations). In this scenario, the systems retain the Ru-O rotation from the zero-strain configuration.  
Notably, removing the Ru-O distortions primarily affects the band gap.\\



The main text focused on the absolute energy difference between majority and minority bands. However, in Fig.~\ref{fig:S7}, we instead plot the signed difference (spin-up minus spin-down) to highlight how the splitting changes sign across the Brillouin zone. For instance, the splitting inverts going from $\mathrm{R'}\to\Gamma$ to $\Gamma\to\mathrm{R}$ and from $\mathrm{Y'}\to\Gamma$ to $\Gamma\to\mathrm{Y}$. Such sign reversals showcase the complementary nature of the $k$-dependent spin splitting that emerges in the altermagnetic state. 
While taking the absolute value is helpful to quantify the splitting strength, plotting the raw sign clarifies how the splitting evolves---and sometimes flips---across distinct points in $\mathbf{k}$-space.\\

Figure~\ref{fig:S7} shows the resulting $\langle \Delta E(\mathbf{k})\rangle$ along the entire $k$-path, defined by 
\begin{equation}
\langle \Delta E(\mathbf{k})\rangle \;=\; \sum_{j=1}^{N_{\text{occ}}}\!
 \frac{1}{N_k}\sum_{i=1}^{N_k} 
   \bigl[E_{\text{down}}(\mathbf{k}_{i,j}) \;-\; E_{\text{up}}(\mathbf{k}_{i,j})\bigr] \,,
\end{equation}

where $N{k}$ is the number of bands at each $k$-point within the specified energy range, and $N_{\text{occ}}$ is the number of occupied bands. These plots show that $\langle \Delta E(\mathbf{k}) \rangle$ is sensitive to the applied strain. Under compressive strain, the overall magnitude of $\langle \Delta E(\mathbf{k}) \rangle$ decreases compared to the unstrained system. In contrast, under tensile strain, it increases, underscoring how the electronic structure and spin splitting respond to lattice deformations.

\begin{figure*}[h]
\centering
\includegraphics[width=1\textwidth]{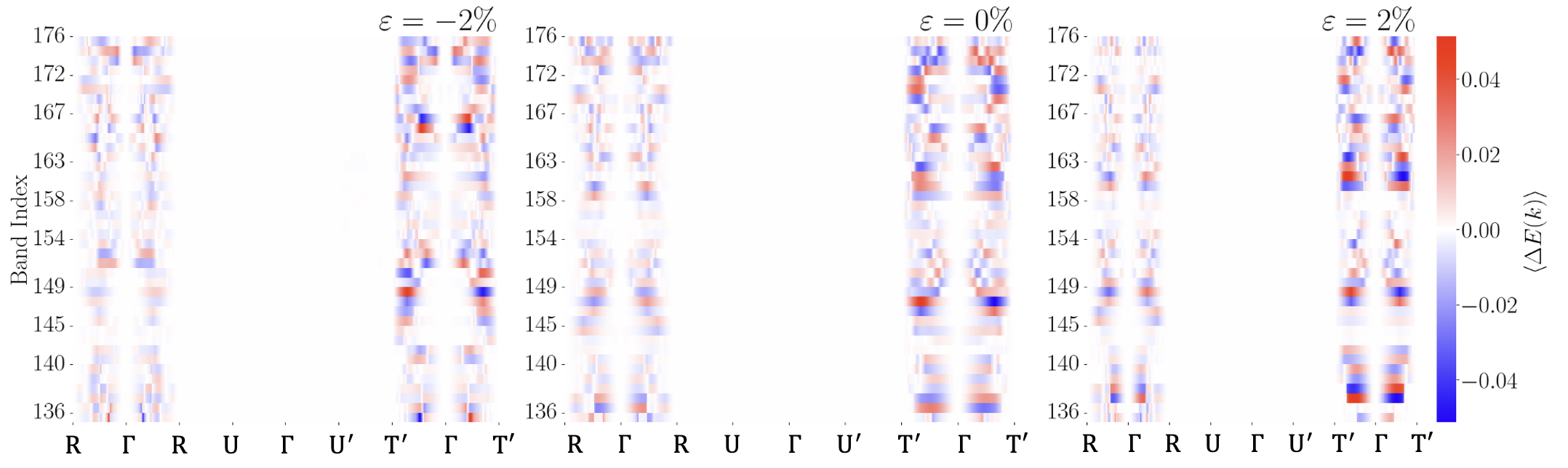}
\caption{Heat maps of $\Delta E(\mathbf{k})$ for the highest occupied bands (where band index 176 corresponds to the higher occupied valence band) along the $
T$--$\Gamma$--$T'$ path, which corresponds to the momentum region exhibiting the strongest altermagnetic splitting (cf.\ the Brillouin zone in Fig.~5).}
\label{fig:S6} 
\end{figure*}

\begin{figure*}[h]
\centering
 \begin{tikzpicture}
    \node at (2.5, 0.5) {spin up};
    \draw[red, ultra thick] (2,0) -- (3,0);
    \node at (5, 0.5) {spin down};
    \draw[blue, ultra thick] (4.5,0) -- (5.5,0);
\end{tikzpicture}\\
\includegraphics[width=0.32\textwidth]{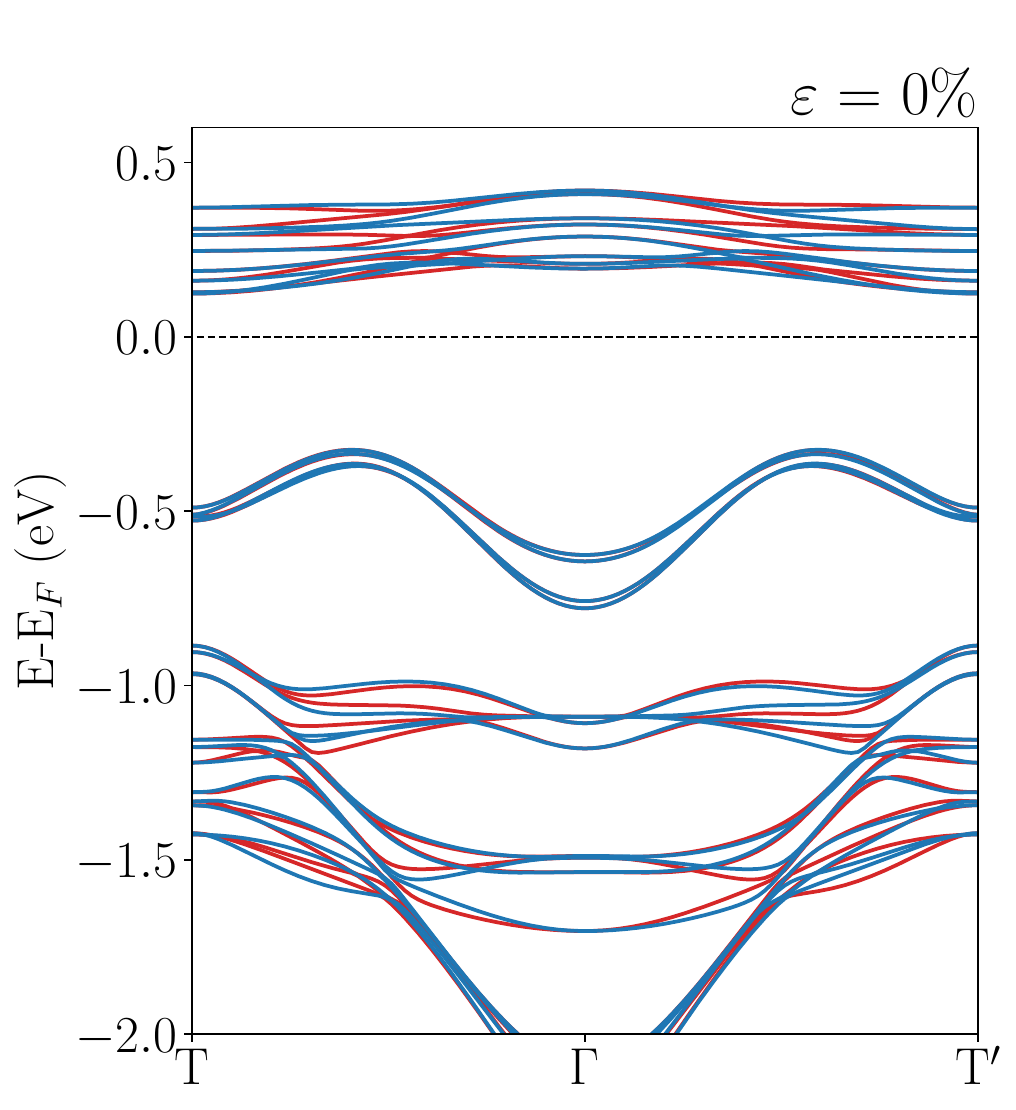}\\
\includegraphics[width=0.95\textwidth]{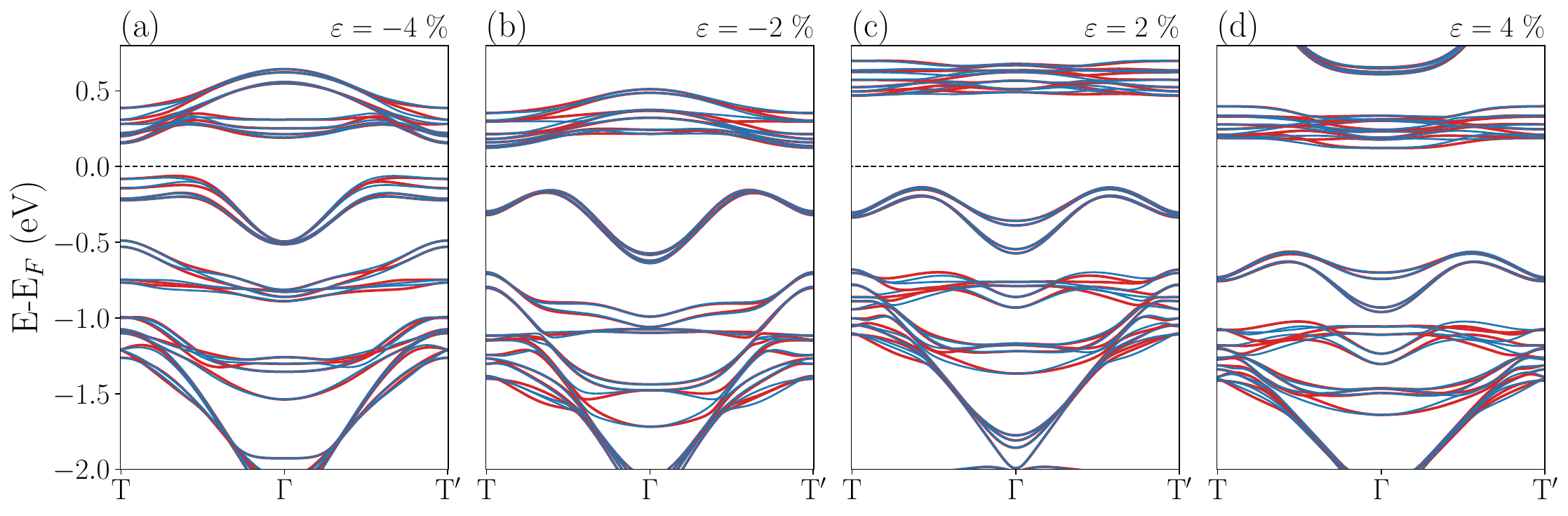}
\includegraphics[width=0.95\textwidth]{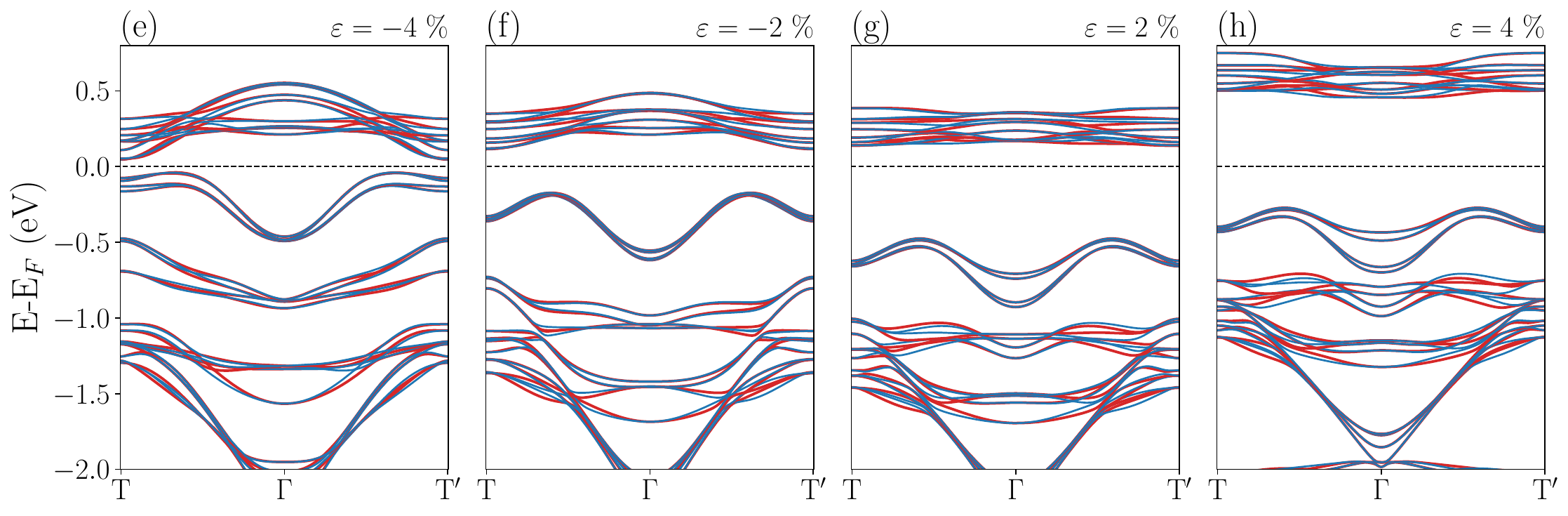}
\caption{(a)-(d) 
Electronic band structure for configuration B under -2\% and -1\% compressive strain, and 1\% and 2\% tensile strain, considering fully relaxed structures. Panels (e)-(h) show the same systems under the corresponding strain conditions, but without ionic relaxation. In these cases, the RuO distortions are kept fixed as in the unstrained structure.
}
\label{fig:S7}
\end{figure*}

\clearpage
\section{3D Multipolar Analysis of the Spin-Resolved Splitting}
 
\subsection*{Conventions and Brillouin-zone averaging}
Let $\mathbf{A}$ be the $3\times3$ matrix of direct-lattice vectors (in \AA) and $\mathbf{B}=2\pi(\mathbf{A}^{-1})^{\mathsf T}$ the reciprocal-lattice matrix (in \AA$^{-1}$). Fractional momenta $\mathbf{k}_{\mathrm{frac}}=(k_1,k_2,k_3)$ are mapped to Cartesian reciprocal coordinates by $\mathbf{k}=\mathbf{B}\,\mathbf{k}_{\mathrm{frac}}$.
Brillouin-zone (BZ) averages are computed discretely using the $k$-point integration weights $w_{\mathbf{k}}$ supplied by the electronic-structure calculation:
\begin{equation}
\big\langle f \big\rangle \equiv
\frac{\sum_{\mathbf{k}} w_{\mathbf{k}}\, f(\mathbf{k})}
{\sum_{\mathbf{k}} w_{\mathbf{k}}}\,.
\end{equation}

\section*{Spin-resolved splitting}
In noncollinear calculations with spin–orbit coupling (SOC), Bloch states are spinors and lack a definite spin projection. The spin-resolved splitting for the spin component $\beta\in\{x,y,z\}$ is constructed from pairs of nearby bands $(i,j)$ via a spin-contrast projection,
\begin{align}
\Delta E_{ij}^\beta(\mathbf{k})
&= \big(E_i(\mathbf{k})-E_j(\mathbf{k})\big)\,u_{ij}^\beta(\mathbf{k}),\\
\mathbf{u}_{ij}(\mathbf{k})
&= \frac{\mathbf{S}_i(\mathbf{k})-\mathbf{S}_j(\mathbf{k})}
{\big\lVert \mathbf{S}_i(\mathbf{k})-\mathbf{S}_j(\mathbf{k})\big\rVert+\epsilon}\,,
\end{align}
where $\mathbf{S}_n(\mathbf{k})$ is the band-resolved spin expectation and $\epsilon>0$ regularizes accidental zeros in the spin contrast (values as small as $10^{-10}$ are sufficient in practice).
For a chosen set $\mathcal{P}$ of $N_{\mathrm{pairs}}$ band pairs (e.g., the topmost occupied spinor bands), the effective splitting is the arithmetic mean
\begin{equation}
\Delta E^\beta(\mathbf{k})
= \frac{1}{N_{\mathrm{pairs}}}\sum_{(i,j)\in\mathcal{P}}
\Delta E_{ij}^\beta(\mathbf{k})\,.
\end{equation}
 
In the collinear (no-SOC) case with quantization axis $\eta$, the splitting is nonzero only for $\beta=\eta$:
\begin{equation}
\Delta E^\beta(\mathbf{k}) =
\begin{cases}
\displaystyle \frac{1}{|\mathcal{B}|}\sum_{b\in\mathcal{B}}
\big(E_{b,\uparrow}(\mathbf{k})-E_{b,\downarrow}(\mathbf{k})\big),
& \beta=\eta,\\[0.4em]
0,& \beta\neq\eta,
\end{cases}
\end{equation}
where $\mathcal{B}$ is the set of bands included in the average.
 
\subsection*{Multipolar moments ($p/d/f$)}
The spin-resolved energy texture $\Delta E^\beta(\mathbf{k})$ is analyzed through the first three BZ moments~\cite{fukaya2025}:
\begin{align}
P_{\alpha}{}^{\beta} &= \Big\langle k_\alpha\,\Delta E^\beta(\mathbf{k}) \Big\rangle,\\
Q_{\alpha\gamma}{}^{\beta} &= \Big\langle k_\alpha k_\gamma\,\Delta E^\beta(\mathbf{k}) \Big\rangle,\\
F_{\alpha\gamma\delta}{}^{\beta} &= \Big\langle k_\alpha k_\gamma k_\delta\,\Delta E^\beta(\mathbf{k}) \Big\rangle,
\qquad \alpha,\gamma,\delta\in\{x,y,z\}.
\end{align}
By parity, $P$ and $F$ are odd under $\mathbf{k}\!\to\!-\mathbf{k}$ and diagnose odd-parity ($p,f$) components, whereas $Q$ is even and diagnoses the even-parity ($d$) component. All averages are evaluated on the full 3D mesh.
 
\subsection*{Strength measures and fractional weights}
For each spin channel $\beta$, the overall strength of $p$-, $d$-, and $f$-wave content is quantified by the $L^1$ norms
\begin{equation}
M_p^\beta=\sum_\alpha \big|P_\alpha{}^\beta\big|,\quad
M_d^\beta=\sum_{\alpha,\gamma}\big|Q_{\alpha\gamma}{}^\beta\big|,\quad
M_f^\beta=\sum_{\alpha,\gamma,\delta}\big|F_{\alpha\gamma\delta}{}^\beta\big|.
\end{equation}
The corresponding fractional weights are
\begin{equation}
f_m^\beta = \frac{M_m^\beta}{M_p^\beta+M_d^\beta+M_f^\beta},
\qquad m\in\{p,d,f\},
\end{equation}
and summarize the dominant multipolar character in each spin channel. It is often convenient to report both per-channel fractions $f_m^\beta$ and overall fractions
$f_m=\sum_\beta M_m^\beta\big/\sum_\beta (M_p^\beta+M_d^\beta+M_f^\beta)$ \cite{fukaya2025,hayami2020}.

\end{document}